\documentclass[aps,10pt,reprint,superscriptaddress,showpacs]{revtex4-2}

\usepackage{xcolor}
\usepackage{physics}
\usepackage{amssymb}
\usepackage{graphicx}
\usepackage{mathtools}
\usepackage{amsmath}
\usepackage{hyperref}
\usepackage{tikz}
\usepackage{amsfonts}
\usepackage{centernot}
\usepackage{dsfont}

\newcommand\equalhat{%
\let\savearraystretch\arraystretch
\renewcommand\arraystretch{0.3}
\begin{array}{c}
\stretchto{
    \scalerel*[\widthof{=}]{\wedge}
    {\rule{1ex}{3ex}}%
}{0.5ex}\\ 
=%
\end{array}
\let\arraystretch\savearraystretch
}

\hypersetup{
colorlinks = true,
urlcolor = blue,
linkcolor = blue,
citecolor = blue
}

\newcommand{\TUM}{\affiliation{Technical University of Munich, TUM School of Natural Sciences, Physics Department, Lichtenbergstr. 4,
85748 Garching, Germany}}
\newcommand{\MCQST}{\affiliation{Munich Center for Quantum Science and Technology (MCQST), Schellingstr. 4, 80799 M{\"u}nchen, Germany}}
\newcommand{\Nottingham}{\affiliation{School of Physics and Astronomy, University of Nottingham, Nottingham, NG7 2RD, UK}}
\newcommand{\CQNE}{\affiliation{Centre for the Mathematics and Theoretical Physics of Quantum Non-Equilibrium Systems, University of Nottingham, Nottingham, NG7 2RD, UK}}
\newcommand{\Oxford}{\affiliation{Rudolf Peierls Centre for Theoretical Physics, University of Oxford, Oxford, UK}}

\begin{document}
\author{Ra{\'u}l Morral-Yepes} \TUM \MCQST 
\author{Adam Smith} \Nottingham \CQNE
\author{S. L. Sondhi} \Oxford
\author{Frank Pollmann} \TUM \MCQST

\title{Entanglement Transitions in Unitary Circuit Games}

\begin{abstract}
Repeated projective measurements in unitary circuits can lead to an entanglement phase transition as the measurement rate is tuned. 
In this work, we consider a different setting in which the projective measurements are replaced by dynamically chosen unitary gates that minimize the entanglement.
This can be seen as a one-dimensional unitary circuit game in which two players get to place unitary gates on randomly assigned bonds at different rates: 
The ``entangler'' applies a random local unitary gate with the aim of generating extensive (volume law) entanglement. 
The ``disentangler'', based on limited knowledge about the state, chooses a unitary gate to reduce the entanglement entropy on the assigned bond with the goal of limiting to only finite (area law) entanglement.
In order to elucidate the resulting entanglement dynamics, we consider three different scenarios: 
(i) a classical discrete height model, 
(ii) a Clifford circuit, 
and (iii) a general $U(4)$ unitary circuit. 
We find that both the classical and Clifford circuit models exhibit phase transitions as a function of the rate that the disentangler places a gate, which have similar properties that can be understood through a connection to the stochastic Fredkin chain.
In contrast, the entangler always wins when using Haar random unitary gates and we observe extensive, volume law entanglement for all non-zero rates of entangling.

\end{abstract}
\maketitle

\section{Introduction}
Quantum many-body systems out of equilibrium represent a challenging frontier and have been shown to exhibit extremely rich phenomena.
These include, for example, a dynamical phase transition between ergodic and many-body localized phases as a function of disorder strength~\cite{MBL0, MBL01, MBL1,MBL2,MBL3,MBL4}, quantum many-body scars~\cite{scars1,scars1.5,scars2,scars3}, and discrete time crystals that can occur in periodically driven systems~\cite{TC0,TC0.5,TC1,TC2,TC3,TC4,TC5}. 
All the above examples occur in closed quantum systems subject to unitary evolution.

\begin{figure}[!t]
    \centering    
    \includegraphics{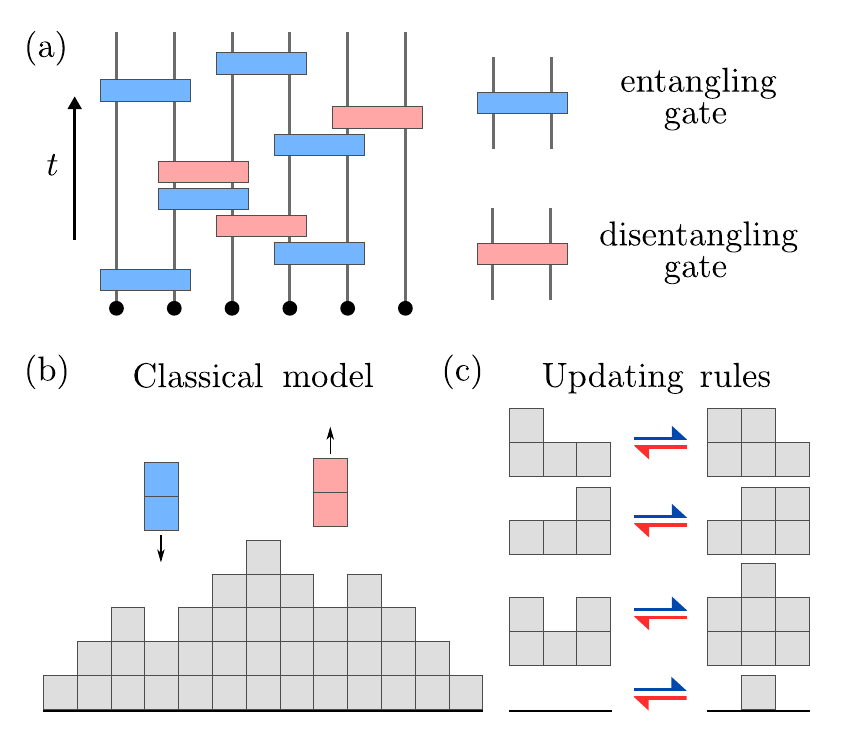}
    
    \caption{(a) Illustration of the unitary circuit game: Blue boxes represent random unitary gates and red gates are unitary gates chosen to disentangle the bond. (b) Classical version of the circuit game as a surface growth model, where entangling and disentangling operations are substituted by adding or removing blocks in a lattice. (c) Updating rules of the classical model: The entangler (blue arrow) adds blocks and the disentangler (red arrow) removes them, with the constraint that the difference of heights between adjacent bonds is at most one and the height cannot be negative.}
    \label{fig: game}
\end{figure}

A new perspective comes from the combination of unitary evolution of a quantum many-body system with measurements.
In pioneering works~\cite{MIPT1,MIPT2,MIPT3,MIPT4, MIPT5}, an entanglement phase transition was identified in the dynamics of circuits of random unitary gates interleaved with local projective measurements.
This phase transition, which separates a disentangling phase obeying an area law from an entangling phase obeying a volume law, has been extensively studied in recent years~\cite{random_circuits_review, MIPT_measurement_only, MIPT_Nahum, A1, A2, A3, A4, A5, A6, A7, A8, A9, MIPT_classical, classicalMIPT2, A11, A12, A13, A14, A15, A16}.
Successively, it has been shown that additional phase transitions between different symmetry broken and topological phases can occur within the area law phase~\cite{MIPTphases1,MIPTphases2,MIPTphases3,MIPTphases4,MIPTphases5,MIPTphases6,10.21468/SciPostPhys.14.3.031}. 

In this paper, we consider a different setting in which the projective measurements are replaced by unitary gates that are dynamically chosen to
disentangle the state. 
While finding the disentangling unitary requires certain knowledge about nonlocal properties of the state, the action of the unitary gate itself is local. 
We can interpret our approach as a (1+1)D circuit game of $L$ sites in which two players get to place unitary gates on randomly assigned bonds at different rates: 
At each updating step, a random bond of the chain is chosen. 
With probability $1-p$ the ``entangler'' acts with a random unitary and otherwise, with probability $p$, the ``disentangler'' acts by minimizing the entanglement entropy on the given bond. 
Here we measure the entanglement entropy $S=-\mathrm{Tr}\rho_A\ln \rho_A$ with the reduced density matrix $\rho_A$, where $A$ includes all qubits left of the bond.
A time step is defined to be a set of $L$ updating steps.
A possible realization of this model is depicted in Fig.~\ref{fig: game}a.
This game has two simple limiting cases: For $p=0$, only the entangler gets to play, resulting in a random-unitary circuit, which in turn leads to a volume law state~\cite{Nahum_entanglement_growth}. 
For $p=1$, the entanglement will remain zero at all times and thus yield an area law state.
Here we are interested in the behavior of our model for intermediate values of $0<p<1$. 
Having defined the rules of the unitary circuit game, several questions naturally emerge:
Is there a phase transition between the volume and area-law entanglement at finite $p$?
If so, what are its universal properties?

We provide answers to these questions in several different variants of the game:
In Sec.~\ref{sec:classical}, we start with a classical surface growth model for which the competition between entangling and disentangling gates is substituted by a competition between increasing and decreasing the height of a surface locally, as shown in Fig.~\ref{fig: game}b. 
Second, in Sec.~\ref{sec:Clifford} we investigate a Clifford circuit. 
In this case, finding the optimal disentangler amounts to selecting from the discrete set of two-qubit Clifford unitary gates.
Finally, in Sec.~\ref{sec:Haar}, we consider general continuously parameterized unitary gates. 
The entangler chooses gates randomly from the Haar distribution, whereas the disentangler now involves the optimization of unitary gates on a given bond in order to minimize the bipartite entanglement entropy.

\section{Classical Model}\label{sec:classical}
In this section, we study a surface growth model in (1+1)D that can be interpreted as a classical version of the unitary circuit game.
We start by introducing the update rules defining the model. 
We then numerically study the phase transition that occurs by tuning the disentangling rate.
Finally, we show that this model is closely related to a stochastic Fredkin spin chain~\cite{FredkinChain}.

The study of entanglement growth in quantum random circuits has gained significant attention in recent years.
In a seminal paper, Nahum et al.~\cite{Nahum_entanglement_growth} showed that the growth of entanglement under random unitary evolution can be mapped to a classical surface growth model, in the limit of infinitely large local Hilbert space dimension. 
The dynamics of this model has the same universal behavior as the entanglement growth in other systems, such as chains of qubits subject to random unitary evolution.
Building on this simplicity, we adopt this model as a starting point and propose a straightforward disentangling rule, essentially undoing the effects of the entangling operations (random unitaries). 
This toy model will serve as a useful benchmark for comparing and contrasting with other more complex quantum models of the unitary circuit game.

For our classical surface growth model, a bond (block) is chosen uniformly at random, and then either the entangler or disentangler takes their go with probability $1-p$ and $p$, respectively.
That is, $p$ controls the disentangling rate. When the entangler is selected, the evolution of the height surface follows the dynamical rule
\begin{equation}\label{classical_entangler}
\begin{aligned}
    &\text{Entangler (probability } 1-p \text{):} \\ & \quad S_x(t+\Delta t) = \min\{S_{x-1}(t), S_{x+1}(t)\} + 1,
\end{aligned}
\end{equation}
where $x$ is the bond index, $t$ denotes the time, and $\Delta t=1/L$.
When the disentangler is selected, the evolution follows the dynamical rule
\begin{equation}\label{classical_disentangler}
\begin{aligned}
    &\text{Disentangler (probability } p \text{):} \\ & \quad S_x(t+\Delta t) = \max\{S_{x-1}(t), S_{x+1}(t), 1\} - 1,
\end{aligned}
\end{equation}
where the one in the argument of the $\max$ function is added to preserve the $S_x(t)\geq0$ constraint for all times, i.e., the height cannot be negative.
We also impose the constraint $S_0(t) = S_{L+1}(t) = 0$, resulting from the open boundary conditions of the model.

The combination of the above rules then defines our classical entangling-disentangling game, as depicted in Fig.~\ref{fig: game}b: at each updating step a random bond is chosen and with probability $p$ the bond is disentangled following Eq.~\eqref{classical_disentangler} or with probability $1-p$ the bond is entangled with Eq.~\eqref{classical_entangler}. 
We can interpret this as a (1+1)D surface growth-depletion model, where the height of each bond corresponds to the entanglement entropy of that bipartition.
The only constraints are that the height cannot be negative and that the difference in height between two adjacent bonds can be at most one, which is the defining condition of a certain class of surface growth models, known as Restricted Solid-On-Solid (RSOS) models~\cite{RSOS, RSOS2,barabasi_stanley_1995}.
The dynamics of certain RSOS models with particle evaporation, similar to our model but with periodic boundary conditions, have been studied in the literature~\cite{deposition_evaporation1, deposition_evaporation2}.

Our classical model was conceived through a direct analogy between the entanglement entropy across a bond and the height of a surface.
Nonetheless, it is worth noting that alternative methods exist for the development of a classical counterpart to our unitary circuit game.
For instance, in Ref.~\cite{MIPT_classical}, an analogy is made between the spread of entanglement due to the random unitary evolution and the spread of information in a classical cellular automaton.
In this particular model, measurements are purely local operations that prevent the spreading of information, mirroring the function of the disentangler in our unitary circuit game.
This alternative framework exhibits a phase transition between a chaotic and a frozen phase, with a critical point with power-law spreading of correlations belonging to the directed percolation universality class.
As we show below, our model presents a similar phase transition with power-law correlations at criticality, but in a different universality class.

\subsection{Numerical results}
\begin{figure}[p]
    \centering
    \includegraphics[width=1\columnwidth]{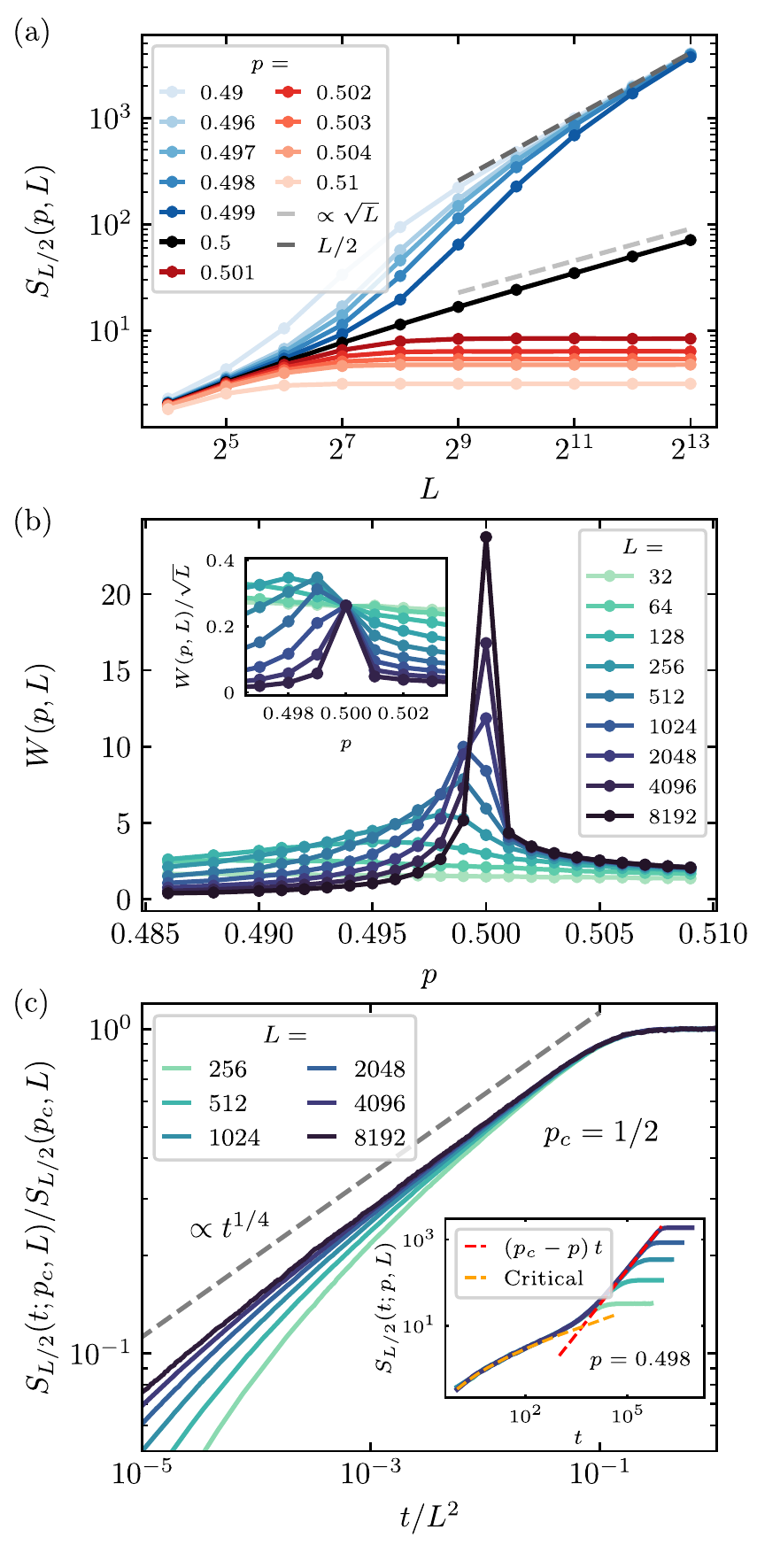}
    \caption{Numerical results for the classical model. (a) Half-chain height as a function of system size $L$ for different values of the disentangling probability $p$. (b) Spatial fluctuations $W(p, L)$ as a function of the disentangling probability $p$ across the phase transition for different system sizes $L$. The inset shows $W(p, L)/\sqrt{L}$, that takes a positive value (independent of system size) at the critical point and tends to zero otherwise. (c) Time evolution, averaged over $10^4$ realizations, of the half-chain height normalized by the steady state value as a function of $t/L^2$ at criticality, $p_c=1/2$. The height increases as a power law with exponent $\beta = 1/4$. The inset shows the time evolution of $S_{L/2}$ at $p=0.498$ (in the volume law phase) for several values of $L$. For $t<t_c$ the evolution is critical (orange line), while for $t>t_c$ the height increases linearly with $t$ (red line).}
    \label{fig: classical S(p)}
\end{figure}

The limiting cases of this model can be easily understood: 
For $p=0$ the disentangler does not act and thus at large times the system reaches a pyramid shaped steady state. 
The increase of height from a flat initial state is described by Kardar-Parisi-Zhang~(KPZ) universal scaling~\cite{KPZ}, as derived in Ref.~\cite{Nahum_entanglement_growth}. 
The limit $p=1$ has a flat steady state, with height 0 at every site. 
Below we investigate the transition between these two limiting cases.

The classical model can be efficiently simulated numerically, allowing us to reach large system sizes, up to $L=8192$. 
For each value of the ``disentangler'' probability $p$ and system size $L$ that we consider, we run $10^3$ different realizations of the circuit (except at the critical point $p_c=1/2$, where we run $10^4$ realizations). 
At each realization, we evolve the system until the dynamical steady state is reached, and then evolve for extra $10^5$ time steps in which we average over the quantities that we are interested in. 
Note that throughout this paper, all quantities are understood to be averaged in the steady state of the system, unless otherwise specified.

Figure~\ref{fig: classical S(p)}a shows the half-chain height $S_{L/2}(p,L)$, as a function of $L$ for several disentangling probabilities $p$ across the phase transition. 
We identify three different behaviors: 
For $p<p_c$ we find a \textit{volume law} phase, where the height increases linearly with system size.
Note that in the thermodynamic limit, all the lines in this phase converge to the line $S_{L/2}=L/2$. 
For $p>p_c$ we find an \textit{area law} phase, where the average height converges to a constant independent of system size. 
Finally, at the critical point $p_c=1/2$ the height is proportional to the square root of system size.
These numerical results strongly suggest a phase transition between volume and area law phases.

To further characterize the different phases, we study the averaged fluctuations around the average steady state height profile of the system using the quantity
\begin{equation}
    W(p,L) = \overline{\left(\frac{1}{L} \sum_{x=1}^{L}\left(S_x(p, L) - \overline{S_x(p, L)}\right)^2\right)^{1/2}},
\end{equation}
where the overline indicates the average of the quantity in the steady state. 
The quantity $S_x(p,L)$ is a stochastic realization of the height profile in the steady state at bond $x$, for disentangling probability $p$ and systems size $L$. 
In practice, we take $t$ much greater than the relaxation time to the steady state, and average over time.
The spatial fluctuations $W(p,L)$ in the classical model are shown in Fig.~\ref{fig: classical S(p)}b as a function of $p$ and $L$. 
In the volume law phase the relative fluctuations tend to zero as the system size is increased, since only in a finite region around the center of the system there will be height variations. 
In the language of stochastic dynamics, this corresponds to an inactive phase.
In the area law phase, fluctuations converge to a constant value for sufficiently large system sizes, indicating an active phase, where fluctuations are equal at every point in the bulk of the system. 
At the critical point, we have a strongly fluctuating phase, where fluctuations increase with system size as $\sqrt{L}$, as shown in the inset of Fig.~\ref{fig: classical S(p)}b.

Finally, we focus on the dynamics of the system and its thermalization time in the critical point $p_c=1/2$.
In particular, we study how the steady state is reached by starting from a flat state with zero height at all sites.
Figure~\ref{fig: classical S(p)}c shows the evolution of the half-chain height averaged over $10^4$ trajectories, normalized by the steady state value. 
The increase of the half-chain height follows a power-law in time, $S_{L/2}(t;p_c, L)\propto t^\beta$, with an exponent $\beta = 1/4$, so that the equilibration time in this phase is $T_{\text{eq}}\propto L^z$, with dynamic exponent $z=2$. 
We observe that the dynamics at the critical point is consistent with Edwards-Wilkinson~(EW) scaling~\cite{EW}, where the exponent $\beta=1/4$ is expected. 
Moreover, we find that the dynamic exponent $z=0$ in the area law phase and $z=1$ in the volume law phase, respectively.
In the inset of Fig.~\ref{fig: classical S(p)}c we show the time evolution of the half-chain entanglement entropy for a disentangling probability in the volume law phase but close to criticality.
We distinguish two regimes: For $t<t_c$, the evolution is the same as in the critical point, i.e., approaching $t^{1/4}$. Instead, for $t>t_c$, the height increases as $(1/2-p)t$, linearly in time. 
The critical time $t_c$, where the evolution changes its behavior, is fixed by the intersection of the two lines. When approaching the critical point it diverges as a power law,
\begin{equation}
    t_c \sim (p_c-p)^{-4/3},
\end{equation}
with critical exponent $4/3$.
In App.~\ref{appendix_dynamics}, we discuss about the dynamics of the system previous to thermalization in more detail, and numerically show that the critical point is governed by the EW universality, while the volume and area law phases follow the KPZ universal scaling.

\subsection{Comparison with Fredkin chain results}

The transition in our classical model can be understood through a connection to the stochastic classical Fredkin chain~\cite{FredkinChain}, a model originally proposed as an example of a power law violation of the area law in quantum spin chains~\cite{MovassaghShor_fredkin,salberger2016fredkin, dellanna_fredkin, Movassagh_fredkin}.
The stochastic Fredkin model is defined on a chain of $L$ sites and each site can be either empty or occupied by a particle ($z_i = 0,1$). 
We focus on the case with $L/2$ particles, i.e., $\sum_i z_i = L/2$. 
This model can be interpreted as a height model by defining the bond variable $h_n = \sum_{i=1}^n (2z_i-1)$, with the condition that $h_n>0$ for every $n$. 
The evolution of the model follows a continuous-time Markov chain evolution, with a parameter $c$ that captures the rate at which transformations occur. 
In particular, the evolution is described by the following updating rules:
\begin{gather*}
1100 \xleftrightharpoons[2c]{2(1-c)} 1010\quad \hat{=} \quad
\begin{tikzpicture}
\filldraw[black] (0,0) circle (1pt);
\draw[black] (0,0) -- (0.2,0.2);
\filldraw[black] (0.2,0.2) circle (1pt);
\draw[black] (0.2,0.2) -- (0.4,0.4);
\filldraw[black] (0.4,0.4) circle (1pt);
\draw[black] (0.4,0.4) -- (0.6,0.2);
\filldraw[black] (0.6,0.2) circle (1pt);
\draw[black] (0.6,0.2) -- (0.8,0.0);
\filldraw[black] (0.8,0.0) circle (1pt);
\end{tikzpicture}
\xleftrightharpoons[2c]{2(1-c)}
\begin{tikzpicture}
\filldraw[black] (0,0) circle (1pt);
\draw[black] (0,0) -- (0.2,0.2);
\filldraw[black] (0.2,0.2) circle (1pt);
\draw[black] (0.2,0.2) -- (0.4,0.0);
\filldraw[black] (0.4,0.0) circle (1pt);
\draw[black] (0.4,0.0) -- (0.6,0.2);
\filldraw[black] (0.6,0.2) circle (1pt);
\draw[black] (0.6,0.2) -- (0.8,0.0);
\filldraw[black] (0.8,0.0) circle (1pt);
\end{tikzpicture},\\ 
    1101 \xleftrightharpoons[c]{\,\,\,\,1-c\,\,\,\,} 1011 \quad \hat{=} \quad 
    \begin{tikzpicture}
\filldraw[black] (0,0) circle (1pt);
\draw[black] (0,0) -- (0.2,0.2);
\filldraw[black] (0.2,0.2) circle (1pt);
\draw[black] (0.2,0.2) -- (0.4,0.4);
\filldraw[black] (0.4,0.4) circle (1pt);
\draw[black] (0.4,0.4) -- (0.6,0.2);
\filldraw[black] (0.6,0.2) circle (1pt);
\draw[black] (0.6,0.2) -- (0.8,0.4);
\filldraw[black] (0.8,0.4) circle (1pt);
\end{tikzpicture}
\xleftrightharpoons[c]{\,\,\,\,1-c\,\,\,\,}
\begin{tikzpicture}
\filldraw[black] (0,0) circle (1pt);
\draw[black] (0,0) -- (0.2,0.2);
\filldraw[black] (0.2,0.2) circle (1pt);
\draw[black] (0.2,0.2) -- (0.4,0.0);
\filldraw[black] (0.4,0.0) circle (1pt);
\draw[black] (0.4,0.0) -- (0.6,0.2);
\filldraw[black] (0.6,0.2) circle (1pt);
\draw[black] (0.6,0.2) -- (0.8,0.4);
\filldraw[black] (0.8,0.4) circle (1pt);
\end{tikzpicture}, \\
    0100 \xleftrightharpoons[c]{\,\,\,\,1-c\,\,\,\,} 0010\quad \hat{=} \quad \begin{tikzpicture}
\filldraw[black] (0,0) circle (1pt);
\draw[black] (0,0) -- (0.2,-0.2);
\filldraw[black] (0.2,-0.2) circle (1pt);
\draw[black] (0.2,-0.2) -- (0.4,0);
\filldraw[black] (0.4,0) circle (1pt);
\draw[black] (0.4,0) -- (0.6,-0.2);
\filldraw[black] (0.6,-0.2) circle (1pt);
\draw[black] (0.6,-0.2) -- (0.8,-0.4);
\filldraw[black] (0.8,-0.4) circle (1pt);
\end{tikzpicture}
\xleftrightharpoons[c]{\,\,\,\,1-c\,\,\,\,}
\begin{tikzpicture}
\filldraw[black] (0,0) circle (1pt);
\draw[black] (0,0) -- (0.2,-0.2);
\filldraw[black] (0.2,-0.2) circle (1pt);
\draw[black] (0.2,-0.2) -- (0.4,-0.4);
\filldraw[black] (0.4,-0.4) circle (1pt);
\draw[black] (0.4,-0.4) -- (0.6,-0.2);
\filldraw[black] (0.6,-0.2) circle (1pt);
\draw[black] (0.6,-0.2) -- (0.8,-0.4);
\filldraw[black] (0.8,-0.4) circle (1pt);
\end{tikzpicture}, \\
    0101 \centernot{\xleftrightharpoons[\hphantom{2c}]{\hphantom{2(1-c)}}} 0011 \quad \hat{=} \quad\begin{tikzpicture}
\filldraw[black] (0,0) circle (1pt);
\draw[black] (0,0) -- (0.2,-0.2);
\filldraw[black] (0.2,-0.2) circle (1pt);
\draw[black] (0.2,-0.2) -- (0.4,0);
\filldraw[black] (0.4,0) circle (1pt);
\draw[black] (0.4,0) -- (0.6,-0.2);
\filldraw[black] (0.6,-0.2) circle (1pt);
\draw[black] (0.6,-0.2) -- (0.8,0);
\filldraw[black] (0.8,0) circle (1pt);
\filldraw[white] (0.4,-0.4) circle (1pt);
\end{tikzpicture}
\centernot{\xleftrightharpoons[\hphantom{2c}]{\hphantom{2(1-c)}}}\begin{tikzpicture}
\filldraw[black] (0,0) circle (1pt);
\draw[black] (0,0) -- (0.2,-0.2);
\filldraw[black] (0.2,-0.2) circle (1pt);
\draw[black] (0.2,-0.2) -- (0.4,-0.4);
\filldraw[black] (0.4,-0.4) circle (1pt);
\draw[black] (0.4,-0.4) -- (0.6,-0.2);
\filldraw[black] (0.6,-0.2) circle (1pt);
\draw[black] (0.6,-0.2) -- (0.8,0);
\filldraw[black] (0.8,0) circle (1pt);
\end{tikzpicture} .
\end{gather*}
Note that these rules avoid the creation of states of the form $0011$, which could lead to negative heights. 

While this model looks similar to the model we have introduced, there are a few key differences:
First, in our classical model, the rate at which the transformations happen is not fixed by a single parameter but instead depends on the state of the system. 
For example, the probability of the disentangler being able to disentangle (instead of doing nothing) is proportional to the number of bonds at which it is possible to reduce the height.
Second, the Fredkin chain dynamics are described by a continuous-time Markov process, in contrast to the discrete dynamics of our model.
Third, the Fredkin model does not allow flat regions, i.e., neighboring sites with the same height.
This is allowed in our model, but they are only created at zero height. These flat regions are irrelevant in the volume law phase in the thermodynamic limit.

While these differences lead to differing dynamics in the two models, the steady states agree at the critical point $c_c = p_c = \frac{1}{2}$. 
Indeed, the difference between continuous and discrete time becomes irrelevant for the average properties of the steady state.
Furthermore, we can understand the critical point as approached from the volume law side where flat regions are irrelevant, and the critical point corresponds to an equilibrium between entangling and disentangling operations.

\begin{figure}[h!]
    \centering
    \includegraphics[width=1\columnwidth]{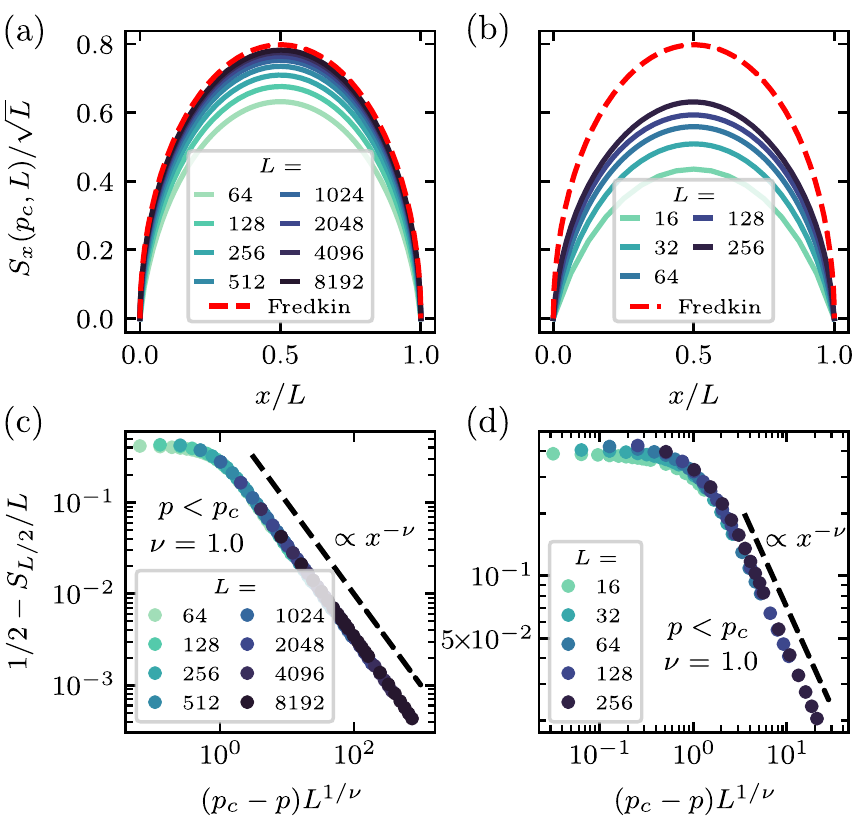}
    \caption{(a) Steady state averaged profile at criticality, $p_c=1/2$, of the classical model for various system sizes, normalized by $\sqrt{L}$. The dashed red line indicates the analytic result for the Fredkin chain in the thermodynamic limit. (b) Steady state averaged profile at criticality, $p_c=0.382$, of the Clifford model. (c) Finite size scaling of the half-chain height normalized by the system size $L$ when approaching the critical point from the volume law phase. The critical exponent $\nu=1$ corresponds to the analytic result in the stochastic Fredkin spin chain. (d) Finite size scaling of the half-chain entanglement entropy in the Clifford model normalized by the system size $L$ when approaching the critical point from the volume law phase, with approximate critical exponent $\nu\approx1.0$.}
    \label{fig: comparison Fredkin}
\end{figure}

The connection to the Fredkin chain is supported by our numerical simulations.
The numerically observed scaling of the height with $\sqrt{L}$ at the critical point agrees with analytical results for the Fredkin chain (given by an average of Dyck paths). 
Additionally, the critical profile of our model converges to the analytical result for the Fredkin chain, which is given by~\cite{Chen_2017_Fredkin}
\begin{equation}
    \label{eq:profile_fredkin}
    \overline{S_x(L)} = \frac{4}{\sqrt{2\pi}}\sqrt{\frac{x(L-x)}{L}},
\end{equation}
(with an extra factor of 2 with respect to the equation in Ref.~\cite{Chen_2017_Fredkin}, since there they consider jumps of 1/2 instead of 1) as depicted in Fig.~\ref{fig: comparison Fredkin}a. 
From a finite size scaling analysis of the half-chain height normalized by the system size, we obtain that the critical exponent for the correlation length when approaching the critical point from the volume law phase is $\nu=1$, as shown in Fig.~\ref{fig: comparison Fredkin}c.
This coincides with the analytical results for the Fredkin chain, where the correlation length scales as~\cite{FredkinChain}
\begin{equation}
    \xi = \ln \left(\frac{c}{1-c}\right)^{-1}\propto |c-1/2|^{-1}.
\end{equation}

\section{Clifford Model}\label{sec:Clifford}

As a first quantum model, we consider unitary circuits that are restricted to Clifford gates \footnote{Clifford gates are unitaries that transform Pauli string operators into Pauli string operators.}.
While simulating general quantum circuits on classical computers is a difficult task, Clifford circuits can be efficiently simulated using stabilizer states~\cite{Cliff1,Cliff2,Cliff3,Cliff4}. 
Clifford circuits offer a sufficiently broad set of operations to show interesting behavior, while keeping the complexity polynomial with system size. 
In fact, Clifford circuits have been the main playground to study measurement-induced phase transitions~\cite{MIPT1,MIPT2,MIPT4}.
The group of Clifford unitaries acting in two qubits is a finite group (containing 11520 unitaries), and therefore finding the optimal unitary to disentangle a bond can be achieved by trying all unitaries until one is able to maximally disentangle the given bond.
Moreover, the entanglement entropy~\footnote{By entanglement entropy, we refer to any Rényi entropy, since for stabilizer states all of them are equal due to the flatness of the entanglement spectrum~\cite{klappenecker2001stabilizer, Winter_stabilizer}.} in the stabilizer states generated by Clifford circuits is always an integer in units of $\log(2)$, and therefore the height picture of the classical model translates to this case, changing the height by the bipartite entanglement. 
However, the rules for entangling and disentangling are more complicated in the Clifford case as compared to the classical model.

\begin{figure}[t!]
    \centering
    \includegraphics[width=1\columnwidth]{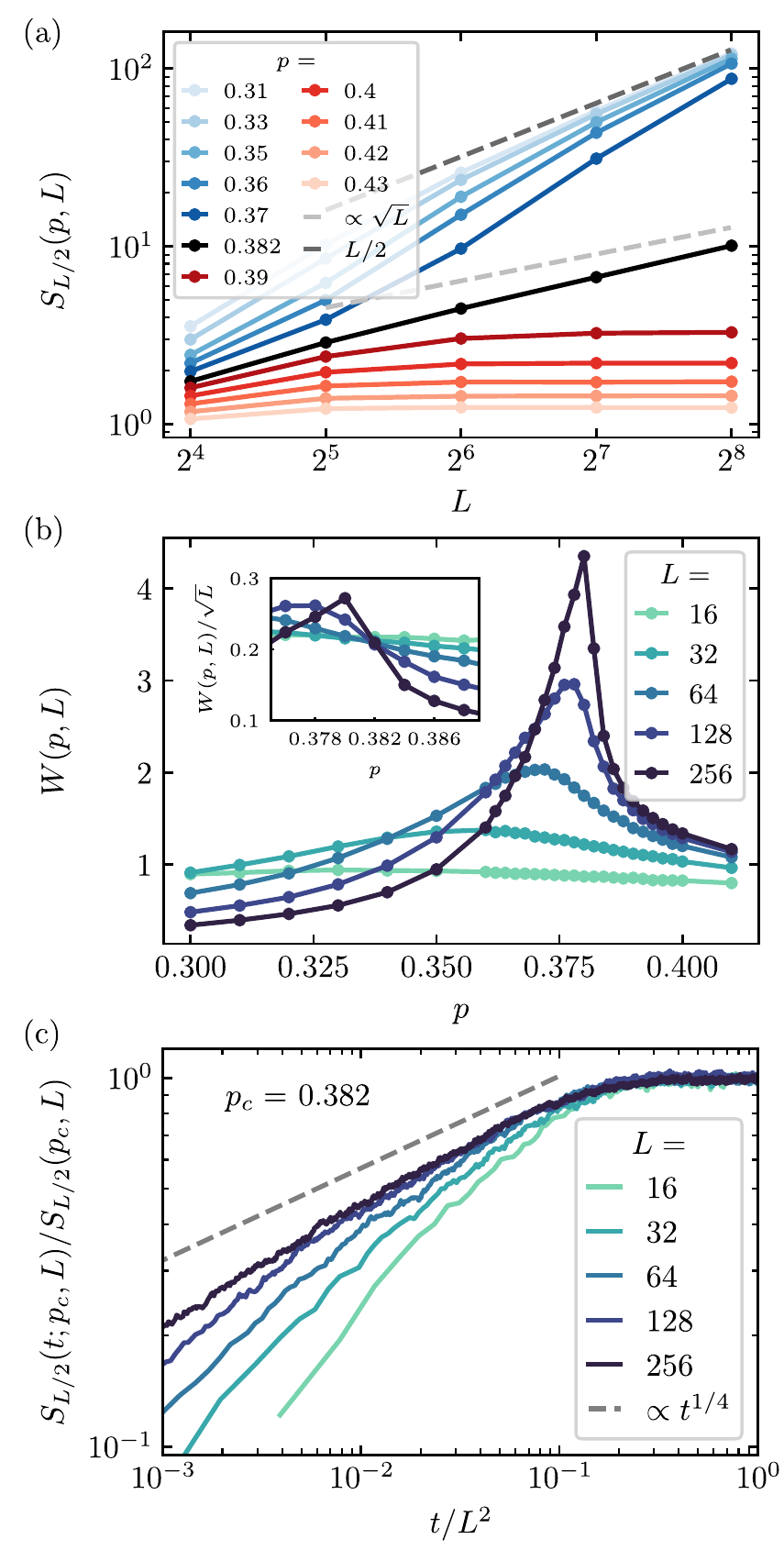}
    \caption{Numerical results for the Clifford model. (a) Half-chain entanglement entropy $S_{L/2}(p, L)$ as a function of system size $L$ for different values of the disentangling probability $p$. (b) Spatial fluctuations $W(p, L)$ as a function of the disentangling probability $p$ across the phase transition for different system sizes $L$. The inset shows $W(p, L)/\sqrt{L}$, which takes a positive value (independent of system size) at the critical point. (c) Time evolution of the half-chain height as a function of time at criticality, $p_c=0.382$, circuit averaged over $10^3$ realizations. The height increases as a power law with exponent $\beta\approx 1/4$, with dynamic exponent $z=2$.}
    \label{fig: clifford S(p)}
\end{figure}

In the Clifford model, we consider a chain of $L$ spin-1/2 degrees of freedom (qubits) with open boundary conditions. 
This chain is evolved as depicted in Fig.~\ref{fig: game}a. 
The entangling gates are drawn randomly and uniformly from the discrete set of Clifford unitary gates, and the disentangling gates are appropriately chosen Clifford unitary gates that maximally reduce the entanglement on that bond.
This model has several key differences compared to the classical one:
First, the entangler in this case is not necessarily optimal. 
Since it applies only a random Clifford unitary, there is a finite probability that it does not increase the entanglement or even that it disentangles the bond. 
Second, the disentangler is not always able to reduce the entanglement as much as it is allowed by rule Eq.~\eqref{classical_disentangler}. 
However, it is always fulfilled that the disentangler at least reduces the entanglement to match that of the adjacent bonds, i.e., $S_x(t+1)\leq \max \{S_{x-1}(t), S_{x+1}(t)\}$.

The disentangling step can be simplified by first looking at the value of the entanglement entropy on adjacent bonds: if $S_x(t)< \min \{S_{x-1}(t), S_{x+1}(t)\}$, then that bond cannot be further disentangled due to subadditivity. 
Otherwise, one has to try unitaries in the Clifford group until one of them is able to maximally disentangle the bond. 
As shown in App.~\ref{appendix_clifford_gates_disentangling}, we find that it is sufficient to choose the disentangling gate from a subset of 19 Clifford unitaries to maximally disentangle any given bond of a stabilizer state.
%
This minimal set of unitaries is not unique, however, the phase transition that we describe in the following is not affected by the choice, see App.~\ref{appendix_clifford_disentanglers} for a discussion about different disentangling methods.

\subsection{Numerical results}

Since the numerical simulations of this model are more demanding than the ones of the classical model, we are limited to system sizes up to $L=256$.
The numerical results show a phase transition between a volume law and an area law phase located at $p_c\approx 0.382$, as shown in Fig.~\ref{fig: clifford S(p)}a. 
The reduction of $p_c$ with respect to the classical model is, in part, caused by the entangler applying random Clifford unitary gates that may not be optimally increasing the entanglement or that could even reduce it. 
For $p<p_c$, the half-chain entanglement entropy asymptotically converges to $L/2$, while for $p>p_c$ it converges to a constant value. 
The behavior at the critical point is similar to the classical case, scaling as $\sqrt{L}$, however, limitations in system size and the large finite size effects (also present in the classical case) do not allow for a precise quantification of the scaling exponent with the numerical data available.
Note that this is in contrast to measurement-induced phase transitions, where the critical point is characterized by a logarithmic scaling of the entanglement entropy and a conformal symmetry of the mutual information~\cite{MIPT4}.

Figure~\ref{fig: clifford S(p)}b shows the behavior of the spatial fluctuations as a function of the disentangler probability. 
The behavior is very similar to the one found for the classical model. 
The divergence of the spatial fluctuations at the critical point allows us to determine with better precision the location of the critical point (inset).
Lastly, we investigate the time evolution of the entanglement entropy in the critical point, as shown in Fig.~\ref{fig: clifford S(p)}c. 
For sufficiently late times, the behavior of the evolution appears to converge to a power law with $t$, with an exponent close to $1/4$ as was the case in the classical model.
App.~\ref{appendix_dynamics} shows additional results for the dynamics of the system, where we find that the critical point is governed by the EW universality class---in agreement with the classical model.
This behavior differs from the critical point of measurement-induced phase transitions, where entanglement entropy grows logarithmically with time and saturates in a time linear with system size~\cite{MIPT4}.
Moreover, we find the same dynamic exponents of $z=2$ in the critical phase, $z=1$ in the volume law phase, and $z=0$ in the area law phase.

The averaged profile in the steady state of the critical point is shown in Fig.~\ref{fig: comparison Fredkin}b. 
The system sizes that can be reached by the numerical simulations are not sufficiently large to convincingly determine whether the profile is converging to one of the Fredkin chain, given by Eq.~\eqref{eq:profile_fredkin}.
However, we can see that even if the proportionality constant is not the same, the profile has qualitatively the same shape as the one in the classical case.
Figure~\ref{fig: comparison Fredkin}d shows the finite size scaling collapse for the half-chain entanglement entropy normalized by the system size.
Based on the limited available data, we find that the critical exponent for the correlation length when approaching the critical point from the volume law is $\nu\approx1$.
Thus the exponent appears to be in agreement with the one found for the classical version of the game.

\section{Haar Random Model}\label{sec:Haar}

\begin{figure}[h!]
    \centering
    \includegraphics{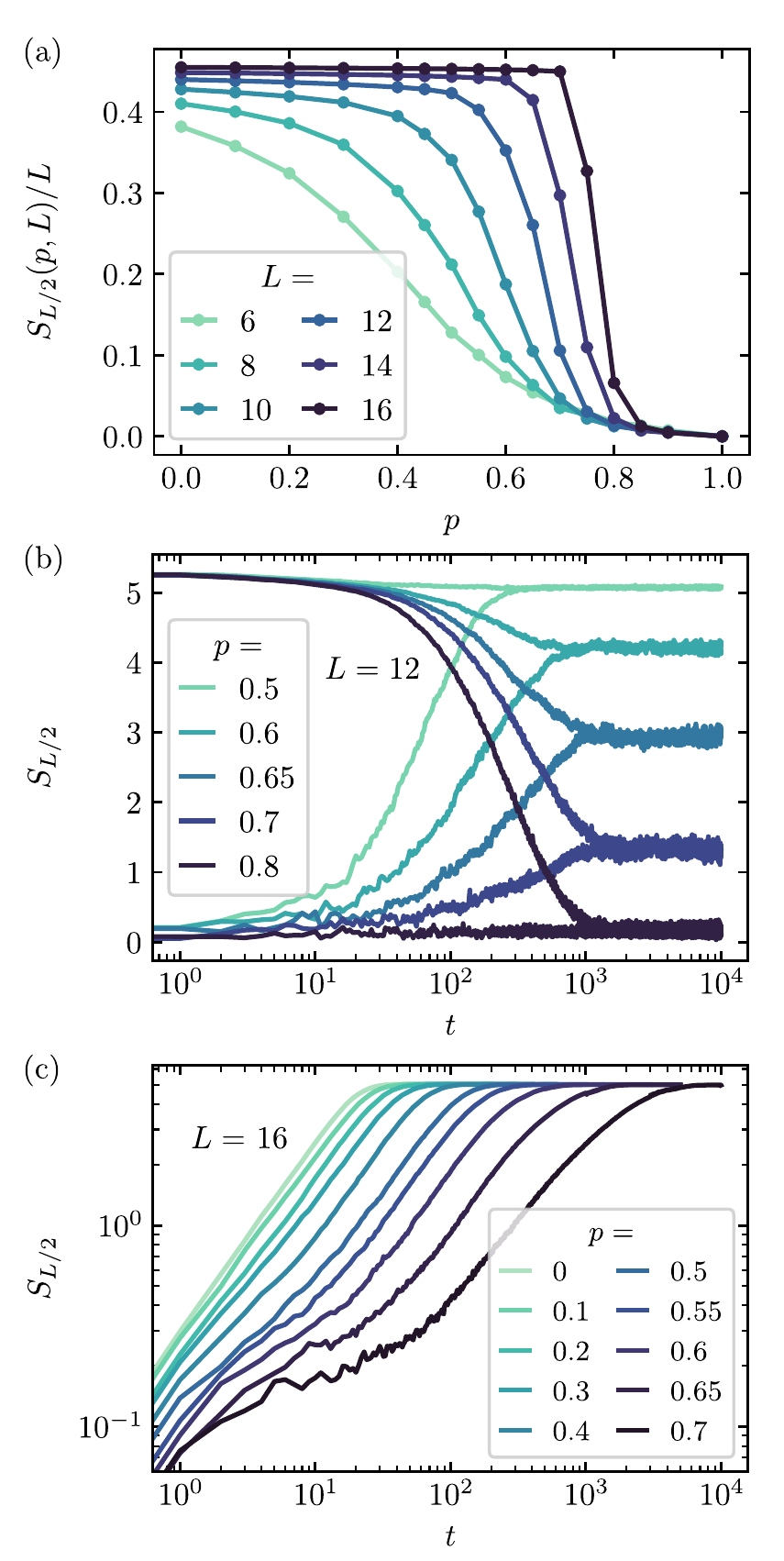}
    \caption{Numerical results for the Haar random model. (a) Order parameter $S_{L/2}/L$ versus the disentangler probability for increasing system size. (b) Time evolution of the half-chain entanglement entropy for fixed system size $L=12$ starting from a product state and from a random Haar state, circuit averaged over 50 realizations. (c) Time evolution for system size $L=16$ and increasing disentangling probabilities. Circuit averaged over 500 realizations until reaching the steady state for short times, and over 10 realizations after that.}
    \label{fig: Haar S(p)}
\end{figure}
We now study the circuit model in the most general case, where unitary gates are taken without restriction from the unitary group $U(4)$. 
While the minimization of the entanglement entropy in the Clifford case was straightforward by choosing a unitary from a discrete set, the disentangling in the Haar case is more challenging.
In particular, we have to find an optimal unitary in $U(4)$ by tuning several continuous parameters---for the case of two site unitary gates, this implies that the disentangler has to perform a minimization with 9 continuous parameters~\cite{entanglement_two-qubit_gates} of a function that is likely to exhibit many local minima.
There exist different entanglement measures that quantify the amount of entanglement of a pure state and thus we have to choose one in order to perform the entanglement minimization.
Here, we will focus on a bipartite von Neumann disentangler, i.e., a disentangler that minimizes the bipartite entanglement entropy $S$ across the given bond~\footnote{We can also consider a R{\'e}nyi disentangler, that minimizes the bipartite $n$-th R{\'e}nyi entropy across the bond. Our numerical results for such disentanglers with $n>1$ show an equivalent behavior to the von Neumann disentangler.}.
In App.~\ref{appendix_disentangling_Haar}, we discuss how the disentangler performs when trying to remove all the entanglement of a state generated by a depth $2L$ random circuit. We find that the disentangling time increases exponentially with system size.

To simulate the system, we rely on the numerically exact evolution of the full wave function of the system and are thus limited to small system sizes, up to $L=16$.

\subsection{Numerical results}

We first investigate the steady-state half-chain entanglement entropy divided by the system size, $S_{L/2}/L$.
This quantity serves as an order parameter for the transition: in a volume law phase, this quantity converges to a finite value in the thermodynamic limit, while in an area law phase it goes to zero. 
Figure \ref{fig: Haar S(p)}a shows this order parameter versus the disentangler probability $p$ for the random Haar circuit with the von Neumann disentangler. 
As the system size is increased, the region in which the system reaches a maximally entangled state is enlarged, indicating that no phase transition exists for any disentangler probability $p<1$.
Therefore, in the thermodynamic limit any effort from the disentangler is futile: even an infinitesimal rate of random unitary gates is expected to eventually lead to a maximally entangled state.
This behavior is very different from the one found in the competition between random unitary evolution and measurements, where there is a phase transition to an area law phase for a sufficiently large, finite rate of measurements.
This relates to the highly nonlocal nature of quantum measurements, in contrast with the local action of our unitary disentangler.

Figure~\ref{fig: Haar S(p)}b shows the circuit averaged time evolution for fixed system size $L=12$ and several different disentangling probabilities, starting from two different initial states: a product state and a Haar random state.
We find that the steady state reached from both possible initial states is the same for any disentangling probability.
Even though the disentangler cannot stabilize an area law phase, it does have an effect in delaying the time required to reach the equilibrium state.
In Fig.~\ref{fig: Haar S(p)}c we fix system size $L=16$ and check the time evolution for increasing disentangler probability $p$.
The time required to achieve the steady state, which is a maximally entangled state, diverges as $p\rightarrow 1$.
In fact, we find that when approaching $p=1$, the equilibration time diverges faster than any power law in $1/(1-p)$ and instead is best described by an exponential divergence.

\begin{figure}

    \centering
    \includegraphics{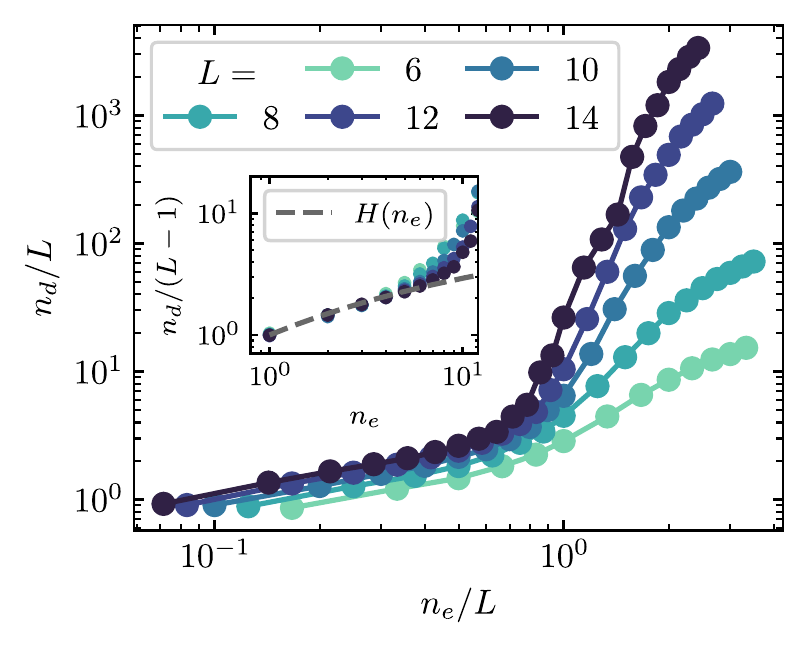}
    \caption{Average number of disentangling steps $n_d$ needed to disentangle a state created by a circuit of $n_e$ randomly placed two-qubit unitary. The condition used to consider the state disentangled is that the sum of the bipartite entanglement at every bond satisfies $\sum_x S_x<10^{-3}L$. The inset shows the collapse for $n_e\ll L$ with $n_d(n_e)/L=H_{n_e}$, with $H_{n}$ the $n$-th harmonic number.}
    \label{fig: disentangle circuit depth}
\end{figure}

We are now going to provide a heuristic argument for the absence of an area law for $p<1$, based on the creation of complex multipartite entanglement structures that the gate-based disentangler is unable to remove effectively.
Let us first consider a simplified model in which we prepare a state by applying $n_e$ random gates to an initial product state.
Next, a disentangling circuit is applied until the total entanglement is reduced below a certain threshold.
Figure~\ref{fig: disentangle circuit depth} shows how the average number of disentangling gates $n_d$ depends on $n_e$.
For $n_e\ll L$, the unitaries applied to the system almost never overlap, such that there is no creation of any multipartite entanglement.
In this case, the disentangler just needs to find the entangled bonds, and then the entanglement can be completely eliminated. 
The average time needed to disentangle can be analytically calculated in this limit, and it is given by $n_d(n_e)=(L-1)H_{n_e}$, where $H_{n_e}$ is the $n_e$-th harmonic number, which coincides with our numerical results (inset Fig.~\ref{fig: disentangle circuit depth}).
Instead, when $n_e \gtrsim L$ the overlapping gates lead to the creation of complex multipartite entanglement structures, making the disentangling task much harder.
For a fixed depth of the entangling circuit, $n_e/L\gtrsim1$, the depth of the disentangling circuit required increases faster than linear with system size.
In particular, as discussed in App.~\ref{appendix_disentangling_Haar}, for $n_e/L = 2L$, the disentangling time grows exponentially in $L$ --- a property that we conjecture to hold for any $n_e/L\gtrsim1$.
In contrast, the circuit depth required by the entangler to create a maximally entangled state grows linearly with system size~\cite{Nahum_entanglement_growth}.
We note that a related setup has been studied in the context of a Metropolis-like entanglement cooling algorithm~\cite{entanglement_cooling}. 
The efficiency of this cooling protocol to remove the entanglement of the state has been related to the complexity of the prepared state and its entanglement spectrum statistics~\cite{entanglement_cooling, entanglement_cooling2, cooling4, cooling3, cooling5}.

Now, we can turn to the unitary circuit game, where the gates that entangle and disentangle occur at a certain rate throughout the time evolution.
For a disentangling rate $p$ close to one, the entangler is initially not able to effectively generate entanglement, since there is a high probability that any random gate will be undone by a disentangling gate before the entangler can place another. 
Nevertheless, given enough time, there is a finite probability that the entangler is able to create a region of length $l$ with multipartite entanglement by overlapping gates.
%
At this point, the time required by the disentangler to remove such a region scales proportional to $\exp(l)$.
Whereas, the probability of the entangler to grow this region to length $l+1$ in a maximally entangled manner scales $\propto l$.
Therefore, given enough time, there is a finite probability that a highly entangled region is created with a sufficient length that it is more likely to grow than shrink.
Such a region will then proliferate and the system will eventually have extensive, volume law entanglement.
In other words, in the thermodynamic limit, the entangler will eventually always win for any $p<1$.
Since the time required to randomly generate an entangled region of length $l$ is exponential in $1/(1-p)$, this argument also explains the exponential in $1/(1-p)$ relaxation time observed in Fig.~\ref{fig: Haar S(p)}c.

\section{Discussion}

Inspired by measurement-induced phase transitions, we introduced a new playground for quantum random circuits in which disentangling measurements are replaced by dynamically chosen unitary gates that minimize the entanglement utilizing limited knowledge about the state.
We investigated three different variants of the model: a classical surface growth model, a Clifford circuit, and a circuit with generic $U(4)$ gates. 
For the classical and Clifford cases, we found a phase transition between a volume law and an area law phase, with a critical point where the entanglement entropy (or height, in the classical case) increases as the square root of the system size. 
We could gain a deeper understanding of the classical model by comparing it to a stochastic Fredkin spin chain.
Regarding the Clifford circuit, we found a qualitative behavior very similar to the classical one, with a phase transition separating a volume law from an area law phase.
However, the numerical limitations in system size did not allow us to determine whether the transition belongs to the same universality class with certainty.
Notably, the behavior of this transition differs significantly from the measurement-induced phase transition observed in Clifford circuits. 
Specifically, the critical point exhibits entanglement growth proportional to the square root of the system size, in contrast to the critical point with logarithmic entanglement observed in measurement-induced transitions.

In the model with random Haar unitaries, we found a qualitatively different behavior: we did not observe a phase transition between volume law and area law for any finite disentangler probability $p<1$.
Instead, we found that the steady state is maximally entangled for any $p<1$ as $L\rightarrow \infty$. 
We provided a heuristic argument for this behavior based on the inefficiency of the disentangler to remove complex structures of multipartite entanglement.
This is something that does not occur in the context of measurement-transitions, where measurements are able to reduce the entanglement irrespective of the complexity of the state, and therefore an area law phase is always observed.

The framework of the unitary circuit game opens many exciting directions for future research in random circuits.
To begin with, going beyond the classical model, it is unclear to which different universality classes the transitions belong to. 
The phase transitions found in the classical and Clifford models have shown to have very similar critical behavior.
However, further investigations are required to determine whether both transitions actually belong to the same universality class.
Another exciting avenue for exploration is to consider different variations of the game, either changing the rules or by restricting the gates to different subsets of $U(4)$---which is expected to lead to different behavior.
Additionally, our current work employs a disentangler that minimizes the entanglement entropy on a bond, but this strategy does not facilitate a transition in the case with generic $U(4)$ unitaries for any finite rate of disentangling.
It remains an open question whether optimizing other quantities could allow for an efficient control of entanglement growth in the thermodynamic limit.
Lastly, the possibility of experimental measurement of R{\'e}nyi entanglement entropies~\cite{experimental2,experimental,experimental3} raises the exciting prospect of implementing the entangling game on physical hardware.

\begin{acknowledgements}
We are grateful to Adam Nahum and Curt von Keyserlingk for insightful discussions. We are grateful to Luke Causer for insightful discussions and for sharing data for the stochastic Fredkin chain. We thank Matteo Ippoliti for helpful comments. R. M. thanks Matthias Englbrecht and Josef Willsher for helpful discussions. This research was financially supported by the European Research Council (ERC) under the European Union’s Horizon 2020 research and innovation program under grant agreement No. 771537. A.S. acknowledges support from a research fellowship from the The Royal Commission for the Exhibition of 1851. This work was supported by a Leverhulme Trust International Professorship grant number LIP-202-014 (S.L.S). For the purpose of Open Access,
the author has applied a CC BY public copyright licence to any Author Accepted Manuscript version arising from this submission. F.P. acknowledges the support of the Deutsche Forschungsgemeinschaft (DFG, German Research Foundation) under Germany’s Excellence Strategy EXC-2111-390814868. F.P.’s research is part of the Munich Quantum Valley, which is supported by the Bavarian state government with funds from the Hightech Agenda Bayern Plus.

\textbf{Data and materials availability:} Raw data and simulation codes are available in Zenodo upon reasonable request~\cite{Zenodo}.
\end{acknowledgements}

\begin{appendix}

\section{Universal dynamics}
\label{appendix_dynamics}

In this paper, we have mainly focused on the averaged steady-state properties of the different models considered. In this Appendix, we are going to study the dynamics of the models prior to thermalization, i.e., how the steady state is reached. We will mainly focus on the classical model since it allows for extensive numerical simulations. We will then show that the Clifford model shows similar behavior up to the system sizes that can be realized.

As discussed in Ref.~\cite{Nahum_entanglement_growth}, the entanglement entropy of our unitary circuit game at $p=0$ grows according to the Kardar-Parisi-Zhang~(KPZ) equation~\cite{KPZ},
\begin{equation}
\label{KPZ}
    \frac{\partial S(x,t)}{\partial t} = \nu \frac{\partial^2 S}{\partial x^2}-\frac{\lambda}{2}\left(\frac{\partial S}{\partial x}\right)^2+\eta(x,t)+c,
\end{equation}
where $\eta(x,t)$ is an uncorrelated noise term and $c$ gives the linear growth behavior. The $c$ term can be absorbed into the height field by substituting $S(x,t)\rightarrow S(x,t)-ct$, thus contributing to the linear growth of entanglement entropy. The $\nu$ is sometimes referred to as the surface tension since it contributes to the smoothing of the interface. Finally, the nonlinear $\lambda$ term describes the dependence of the growth rate on the slope of the surface. For $\lambda=0$, Eq.~\eqref{KPZ} reduces to the Edwards-Wilkinson~(EW) equation~\cite{EW}, which has different universal properties with respect to KPZ.

In this section, we will look at three different exponents that characterize the dynamics~\cite{barabasi_stanley_1995}: the \textit{growth exponent} $\beta$ characterizing the size of the fluctuations in the interface, the \textit{roughness exponent} $\alpha$ characterizing the spatial fluctuations, and the \textit{dynamic exponent} $z_d$ that sets the rate of growth of the correlation length $\xi_{||}$, defined as the characteristic distance of the spatial correlations. Note that $z_d$ is not the same as the dynamic exponent $z$ presented in the main text, which characterizes the equilibration time of the height (or entanglement entropy). Instead, $z_d$ characterizes the saturation time of the fluctuations. The way the exponents relate to the dynamics is the following. The average height increases linearly with a subleading correction,
\begin{equation}
    \overline{S_x(t)}=v_E t + Bt^\beta,
\end{equation}
where the overline indicates an average over trajectories. The fluctuations grow as
\begin{equation}
    \label{eq:fluctuations}
    w_x(t)\equiv\left(\overline{S_x(t)^2}-\overline{S_x(t)}^2\right)^{1/2}=C t^\beta.
\end{equation}
The ratio $C/B$ is a universal quantity. The spatial correlation length growths with time as $\xi_{||}(t)\propto t^{1/z}$ and the spatial correlations fulfill
\begin{equation}
\label{eq:spatial_correlation}
    G(r)\equiv \left[\overline{\left(S_x(t)-S_{x+r}(t)\right)^2}\right]^{1/2}=r^\alpha g\left(r/\xi_{||}(t)\right).
\end{equation}
Note that the spatial correlations are only measured in the ``active'' region of the profile, i.e., far away from the positions where the height has already saturated to a fixed value. In practice, to determine the spatial fluctuations, we fix $x=L/2$, since this is the point that takes longer to reach the steady state.

The universal KPZ scaling behavior is given by the exponents
\begin{equation}
    \text{KPZ:}\qquad\beta = 1/3, \qquad \alpha = 1/2, \qquad z_d=3/2.
\end{equation}
Instead, the EW universality is characterized by
\begin{equation}
    \text{EW:}\qquad\beta = 1/4, \qquad \alpha = 1/2, \qquad z_d=2.
\end{equation}

\subsection{Dynamics of the classical model}

\begin{figure}[t!]
    \centering
    \includegraphics{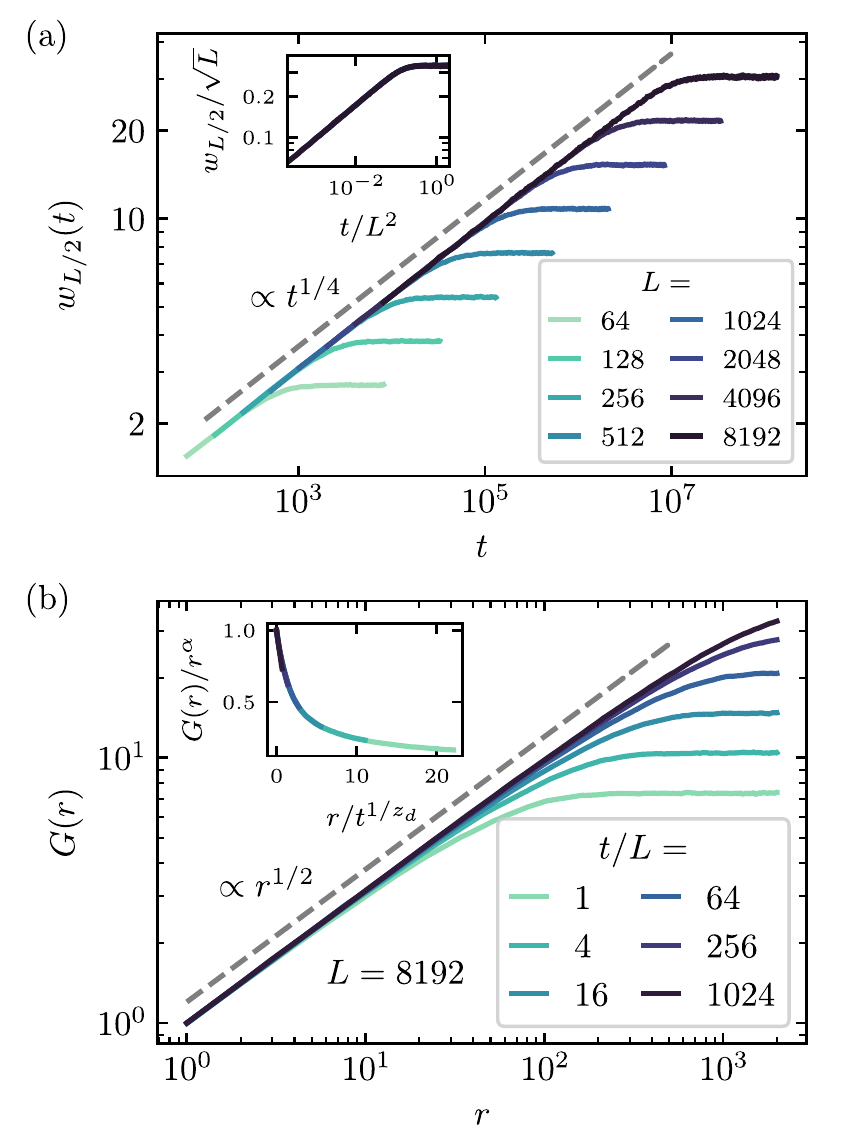}
    \caption{Dynamical properties of the classical model at the critical point $p_c=1/2$. (a) Evolution of the half-chain height fluctuations for increasing system size. Inset: Scaling collapse of the evolution. (b) Spatial correlations as a function of distance $r$ to the center of the chain for system size $L=8192$ at different time steps. Inset: Scaling collapse given by Eq.~\eqref{eq:spatial_correlation} with EW exponents, $\alpha=1/2$ and $z_d=2$. Data averaged over $2\times10^4$ circuit realizations for $L=8192$ and $10^5$ realizations otherwise.}
    \label{fig: classical critical dynamics}
\end{figure}

We start considering the classical model at the critical point $p_c=1/2$.
As we have seen in Fig.~\ref{fig: classical S(p)}c, the velocity at the critical point vanishes, and the height grows as $t^{1/4}$, consistent with EW scaling.
Now, we look at other dynamical properties of the system at criticality.
Fig.~\ref{fig: classical critical dynamics}a shows the evolution of the fluctuations with time in the half-chain.
The growth follows a power law with exponent $\beta=1/4$.
Fig.~\ref{fig: classical critical dynamics}b shows the spatial correlations for fixed system size $L=8192$ at different time steps. As expected, the spatial correlations follow a power law, with exponent $\alpha=1/2$.
The inset of the figure shows the collapse obtained using the scaling form~\eqref{eq:spatial_correlation} with dynamic exponent $z_d=2$.
All these results confirm that the dynamics of the critical point are captured by the EW universality class.

\begin{figure}[t!]
    \centering
    \includegraphics{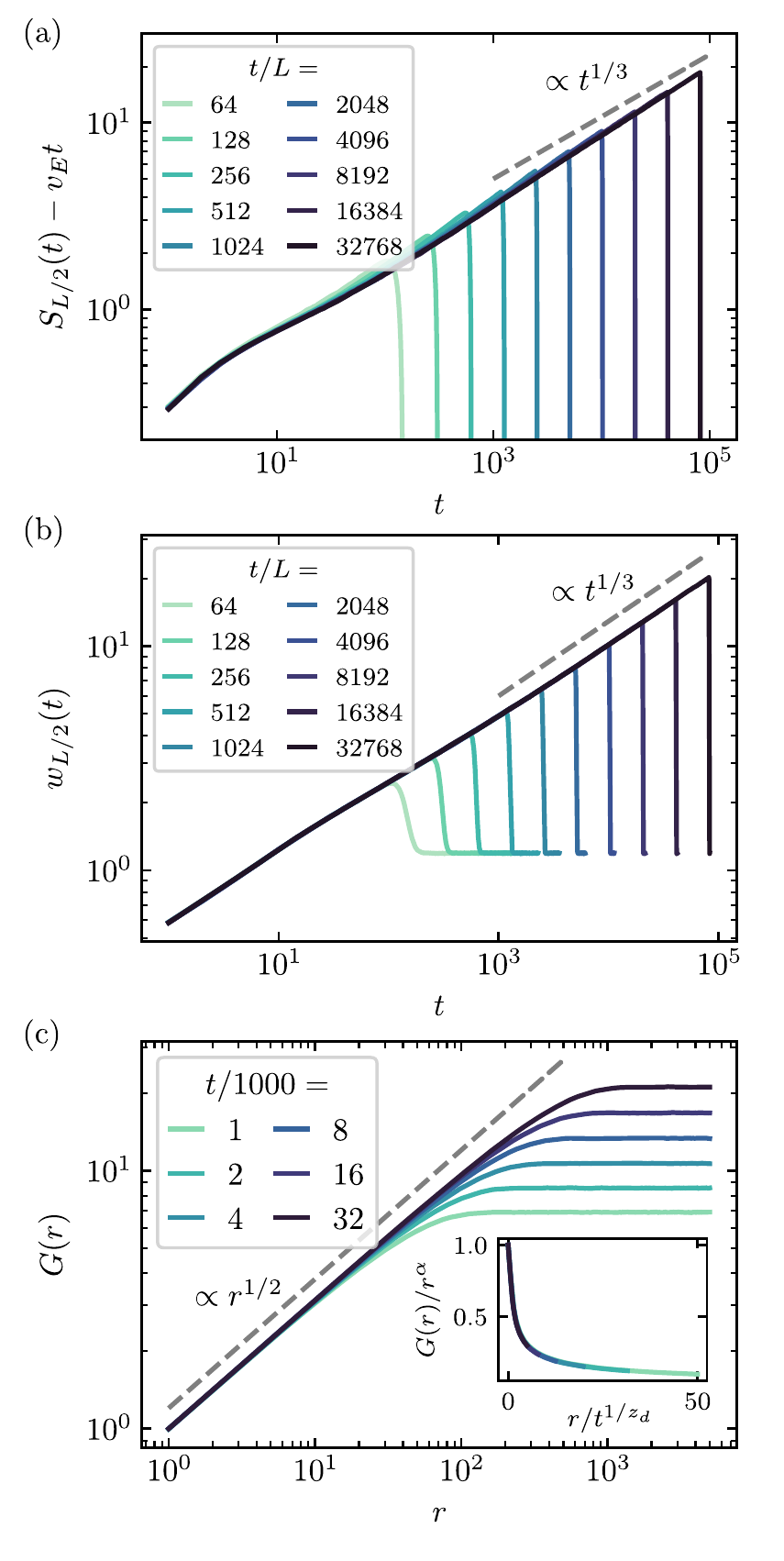}
    \caption{Dynamical properties of the classical model at the volume law phase, for $p=0.3$. (a) Subleading growth of the half-chain height $S_{L/2}(t)-v_Et$ with time for increasing system size. At late times, the growth follows a power law with exponent $\beta = 1/3$ (b) Evolution of the half-chain height fluctuations for increasing system size. (c) Spatial correlations as a function of distance $r$ to the center of the chain for system size $L=32768$ at different time steps. Inset: Scaling collapse given by Eq.~\eqref{eq:spatial_correlation} with KPZ exponents, $\alpha=1/2$ and $z_d=3/2$. Data averaged over $10^5$ circuit realizations.}
    \label{fig: classical volume dynamics}
\end{figure}

We now turn to the dynamics in the volume law phase. As mentioned in the main text, the velocity at which the height grows is given by $v_E=p_c-p$, which vanishes at the critical point. Therefore, to see the action of the nonlinear term of Eq.~\eqref{KPZ} in the growth of entanglement, we need to subtract the linear velocity term. Fig.~\ref{fig: classical volume dynamics} shows the dynamics of the system at the point with disentangling probability $p=0.3$. In Fig.~\ref{fig: classical volume dynamics}a we show the subleading increase of the height coming from the nonlinear term in the KPZ equation, which for long enough times grows as $t^\beta$, with $\beta=1/3$. Similarly, in Fig.~\ref{fig: classical volume dynamics}b we find that the fluctuations grow as predicted by Eq.~\eqref{eq:fluctuations}, with the exponent corresponding to the KPZ universality. We note that in both these cases the KPZ behavior appears only for long times, and this becomes more extreme as we approach the critical point. This is caused by the crossover between EW and KPZ behavior: as the critical point is approached, the nonlinear term controlled by $\lambda$ becomes smaller ($\lambda$ tends to 0 as $p\rightarrow 1/2$), and the time scales at which it becomes relevant increase. In particular, the crossover from EW to KPZ behavior happens for times proportional to $\lambda^{-\phi}$, with $\phi=4$~\cite{deposition_evaporation1, deposition_evaporation2}.

Fig.~\ref{fig: classical volume dynamics}b shows that the fluctuations $w_{L/2}$ grow with time as a power law until reaching a maximum and, at this point, they decrease fast to a fixed value independent of system size. This phenomenon is caused by the open boundary conditions of the system and the subadditivity constraint, which impose a maximum height of $L/2$ in the middle of the chain. Then, when the height becomes large enough to reach the upper boundaries, the fluctuations are reduced. With periodic boundary conditions, there is no upper limit in height, and the fluctuations increase until reaching $w_{L/2}\sim L^\alpha$ in a time $\sim L^{z_d}$.

Finally, Fig.~\ref{fig: classical volume dynamics}c shows the spatial correlations as a function of the distance for several time steps. As predicted by the KPZ scaling, the correlations follow a power law with exponent $\alpha=1/2$. In the inset, we show a finite size scaling collapse following the ansatz~\eqref{eq:spatial_correlation}. The results confirm the dynamic exponent $z_d=3/2$, characteristic of KPZ behavior, in contrast with the behavior at the critical point.

Within the area law phase, the results are symmetric with respect to the volume law phase. However, to see the KPZ universality in this phase, one needs to initialize the system in the state with maximum height. Then, the results are similar to those obtained in the volume law phase, with a velocity $v_E=p_c-p$ that is now negative. The fluctuations in the steady state are limited by the boundary condition $S_x(t)\geq0$, so fluctuations grow as the height is reduced until they reach the lower boundary of the system, showing a similar behavior as in Fig.~\ref{fig: classical volume dynamics}b.

\subsection{Dynamics of the Clifford model}
\begin{figure}[t!]
    \centering
    \includegraphics{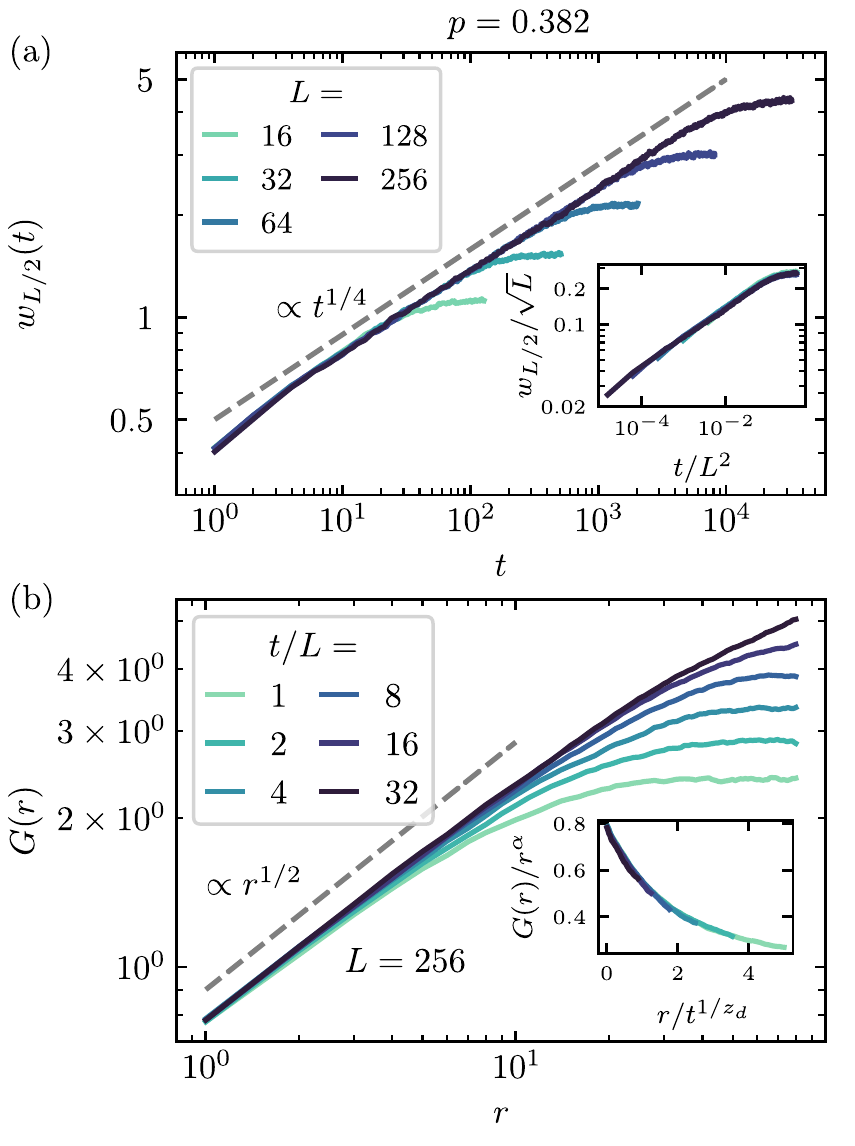}
    \caption{Dynamical properties of the Clifford model at the critical point $p_c=0.382$. (a) Evolution of the half-chain height fluctuations for increasing system size. Inset: Scaling collapse of the evolution. (b) Spatial correlations as a function of distance $r$ to the center of the chain for system size $L=256$ at different time steps. Inset: Scaling collapse given by Eq.~\eqref{eq:spatial_correlation} with EW exponents, $\alpha=1/2$ and $z_d=2$. Data averaged over $10^4$ circuit realizations.}
    \label{fig: clifford critical dynamics}
\end{figure}

In the Clifford model, we obtain similar results as for the classical one. In Fig.~\ref{fig: clifford critical dynamics} we show the dynamical properties of the system at the critical point $p_c=0.382$. We find equivalent results to those obtained in Fig.~\ref{fig: classical critical dynamics} for the classical model. In particular, the fluctuations increase as a power law with exponent $\beta=1/4$ and the correlations increase with distance with $\alpha=1/2$, as predicted by the EW scaling. 

Analyzing the results in the volume law phase in the Clifford model is more challenging due to the limitations in system size and the long time scales at which the crossover from EW to KPZ happens. Thus, we leave a detailed study of the dynamics at such phases for future work.

\section{Minimal set of disentangling two-qubit Clifford unitaries}
\label{appendix_clifford_gates_disentangling}

In this Appendix, we demonstrate that it is possible to disentangle any bond within a stabilizer state using a Clifford unitary selected from a minimal set of 19 two-qubit unitaries. Our proof will take a constructive approach, providing the explicit form of these unitaries. The organization of this Appendix is as follows: First, we will provide a brief summary of the stabilizer formalism, which is utilized for simulating Clifford circuits. We review the concept of the clipped gauge, which is used to express the system's state in a manner that facilitates direct computation of entanglement entropies. Finally, we employ the clipped gauge to identify all local entanglement structures and determine the unitaries capable of disentangling them.

\subsection{Review of the stabilizer formalism}

A stabilizer state of $L$ qubits is defined to be a quantum state $\ket{\psi}$ for which there are $L$ linearly independent and mutually commuting Pauli string operators $g_i$ that leave the state invariant, i.e., $g_i\ket{\psi}=\ket{\psi}$. Such operators are called stabilizers. The group generated by these stabilizers and matrix multiplication is known as the stabilizer group of the state $\ket{\psi}$.

A Clifford unitary is a unitary that maps every Pauli string into another Pauli string, therefore preserving stabilizer states. It is a well-known fact that the Clifford group is generated by the set of gates $\{\text{CNOT}, \text{S}, \text{H}\}$. The number of Clifford gates is finite since there is a finite number of Pauli strings. In particular, for two-qubit unitaries, it consists of 11520 gates. 

The Gottesmann-Knill theorem~\cite{Cliff1,Cliff2,Cliff3,Cliff4} ensures that any circuit consisting of an initial stabilizer state evolved with Clifford gates and Pauli measurements can be efficiently simulated in a classical computer.

The entanglement entropy of a stabilizer state in a region $A$ can be calculated by counting the number of stabilizers in the stabilizer group that are completely contained in $A$, i.e., that act trivially in its complementary. Then, the entanglement entropy is given by~\cite{stabilizer_entanglement}
\begin{equation}
    S_A = n_A-\log_2|\mathcal{S}_A|,
\end{equation}
where $n_A$ is the number of qubits contained in $A$ and $|\mathcal{S}_A|$ is the size of the subgroup of stabilizers contained in $A$.

We observe that the set of stabilizers generating the stabilizer group can be chosen in many different ways, since the product of any two stabilizers is still a stabilizer. This is referred to in the literature as gauge freedom~\cite{MIPT4}. A particularly useful gauge is known as the clipped gauge for one-dimensional spin chains, first introduced in Ref.~\cite{Nahum_entanglement_growth}. Define $\rho_l(i)$ to be the number of stabilizers with the left endpoint (i.e., the first site with a nontrivial content) in site $i$, and equivalently for $\rho_r(i)$ with right endpoints. Then, the generating set of stabilizers can be always chosen in such a way that the following properties are fulfilled:
\begin{itemize}
    \item $\rho_r(i)+\rho_l(i)=2$ for every site $i$.
    \item If a site $i$ has $\rho_r(i)=2$ or $\rho_l(i)=2$, then the two stabilizers have a different Pauli operator in site $i$.
\end{itemize}
It is always possible to bring a stabilizer state to the clipped gauge, for an algorithm see Ref.~\cite{MIPT4}. A state in the clipped gauge can be diagrammatically represented in the following way~\cite{MIPTphases1}: draw $L$ points representing the sites. Then, for each stabilizer, draw a line connecting the left endpoint to the right endpoint. For example, a stabilizer state of 8 qubits could have the following representation:
\begin{equation}
\label{eq:example state}
\begin{tikzpicture}[scale=1]
\draw[blue] (0,0) .. controls (1,1) and (2,1) .. (3,0);
\draw[blue] (0,0) .. controls (1,-1) and (2,-1) .. (3,0);
\draw[blue] (1,0) .. controls (0.6,0.6) and (1.4,0.6) .. (1,0);
\draw[blue] (2,0) .. controls (3,1) and (4,1) .. (5,0);
\draw[blue] (2,0) .. controls (2.66,-0.8) and (3.33,-0.8) .. (4,0);
\draw[blue] (4,0) .. controls (4.66,-0.8) and (5.33,-0.8) .. (6,0);
\draw[blue] (5,0) .. controls (5.66,0.8) and (6.33,0.8) .. (7,0);
\draw[blue] (6,0) .. controls (6.33,-0.5) and (6.66,-0.5) .. (7,0);
\filldraw[black] (0,0) circle (2pt);
\filldraw[black] (1,0) circle (2pt);
\filldraw[black] (2,0) circle (2pt);
\filldraw[black] (3,0) circle (2pt);
\filldraw[black] (4,0) circle (2pt);
\filldraw[black] (5,0) circle (2pt);
\filldraw[black] (6,0) circle (2pt);
\filldraw[black] (7,0) circle (2pt);
\end{tikzpicture}
\end{equation}
Note that this representation does not uniquely represent a stabilizer state. Nevertheless, it contains all the information that we need about the entanglement entropy.

In the clipped gauge, the entanglement entropy of a contiguous region $A$ is particularly simple to calculate: it is given by the number of generators with an endpoint inside $A$ and the other outside $A$. Therefore, the bipartite entanglement entropy can be directly obtained from the diagrammatic representation of the state by looking at how many strings cross each bond. For example, the entanglement profile in the example state~\eqref{eq:example state} would be $S(x)=(1,1,2,1,1,1,1)$. Using this representation of stabilizer states, we observe that the entanglement can be reduced locally only by moving the endpoints of the stabilizers within the qubits in which the unitary acts. In the following section, we are going to identify all the local entanglement structures that one can have in a given bond and find a way to maximally disentangle each of them.

\subsection{Locally disentangling a state}

To disentangle a bond in a stabilizer state, one needs, in principle, to find the optimal disentangling unitary among the set of two-qubit Clifford unitaries. However, this set contains many unitaries, and trying out all of them is a very time-consuming task. Nevertheless, within stabilizer states the allowed local structures of entanglement are limited. In this section, we will explore all potential local entanglement structures and the corresponding unitaries required for disentangling them. We will find that a finite set of 19 two-qubit Clifford unitaries is enough to maximally disentangle any possible bond.

The first step to finding the disentangling Clifford unitaries is to look at the possible local stabilizer structures, this is, all possible configurations of stabilizer endpoints in the clipped gauge in adjacent qubits. By counting in how many ways one can add strings with two endpoints in each site, we determine that there are 21 possible stabilizer structures (note that each of them corresponds, at the same time, to many local states). Only some of these structures can be disentangled. For example, consider the following two local stabilizer structures:
\begin{equation}
    \begin{tikzpicture}[scale=1]
        \clip (-0.9,0.85) rectangle + (2.2,-1.3);
        \draw[blue] (0,0) .. controls (-1,0.5) and (-2,0.5) .. (-3,0);
        \draw[blue] (0,0) .. controls (-1,0.8) and (-2,0.8) .. (-3,0);
        \draw[blue] (1,0) .. controls (0.6,0.8) and (1.4,0.8) .. (1,0);
        \filldraw[black] (0,0) circle (2pt);
        \filldraw[black] (1,0) circle (2pt);
    \end{tikzpicture}
    \qquad \quad 
    \begin{tikzpicture}[scale=1]
        \draw[gray] (0,00)  -- (0,+1);
    \end{tikzpicture} 
    \quad\qquad
    \begin{tikzpicture}[scale=1]
        \clip (-0.6,0.85) rectangle + (2.2,-1.3);
        \draw[blue] (1,0) .. controls (1.66,0.6) and (2.33,0.6) .. (3,0);
        \draw[blue] (0,0) .. controls (-0.66,0.6) and (-1.33,0.6) .. (-2,0);
        \draw[blue] (0,0) .. controls (0.33,-0.5) and (0.66,-0.5) .. (1,0);
        \filldraw[black] (0,0) circle (2pt);
        \filldraw[black] (1,0) circle (2pt);
    \end{tikzpicture}
\end{equation}
In the first case, the stabilizers whose endpoints are modified by a local unitary do not cross the bond, and therefore the entanglement entropy cannot be reduced locally. In the second case, there is a stabilizer that crosses the bond. However, there is no way to move the endpoints of the stabilizers in such a way that there is less than one string crossing it while keeping two endpoints per site. Therefore, the configuration cannot be further disentangled with a local unitary. All such configurations are said to be \textit{locally minimal}. There are 11 such configurations. The other 10 configurations are denoted \textit{locally entangled} and can be, at least in principle, locally disentangled.

Now, let us look at the locally entangled configurations and construct the unitaries that disentangle them. We start looking at the following configurations:
\begin{equation}
    \begin{tikzpicture}[scale=1]
        \clip (-0.6,0.65) rectangle + (2.2,-1.3);
        \draw[blue] (1,0) .. controls (0.6,0.8) and (1.4,0.8) .. (1,0);
        \draw[blue] (0,0) .. controls (1,-0.5) and (2,-0.5) .. (3,0);
        \draw[blue] (0,0) .. controls (1,-0.8) and (2,-0.8) .. (3,0);
        \filldraw[black] (0,0) circle (2pt);
        \filldraw[black] (1,0) circle (2pt);
    \end{tikzpicture}
    \qquad \quad 
    \begin{tikzpicture}[scale=1]
        \draw[gray] (0,00)  -- (0,+1.3);
    \end{tikzpicture} 
    \quad\qquad
    \begin{tikzpicture}[scale=1]
        \clip (-0.6,0.65) rectangle + (2.2,-1.3);
        \draw[blue] (0,0) .. controls (-0.4,0.8) and (0.4,0.8) .. (0,0);
        \draw[blue] (1,0) .. controls (0,-0.5) and (-1,-0.5) .. (-2,0);
        \draw[blue] (1,0) .. controls (0,-0.8) and (-1,-0.8) .. (-2,0);
        \filldraw[black] (0,0) circle (2pt);
        \filldraw[black] (1,0) circle (2pt);
    \end{tikzpicture}
\end{equation}
A loop in a single site corresponds to having a stabilizer completely contained in the site, meaning that the state is invariant under the application of a Pauli operator $P$ in that site (maybe up to a sign). Applying a SWAP gate, which is a Clifford unitary, will move the single-site stabilizer from one site to the other. Note that, in the site with a single stabilizer, the other two stabilizers might only have $P$ or an identity, since those stabilizers must commute with $P$. If they have an identity, then after applying the SWAP gate the stabilizer will automatically start/end in the other site. If they have the same content as the single stabilizer, then they can be multiplied by $P$ (since the product of stabilizers is a stabilizer) to create an identity in that site and reduce it to the first case. Therefore, the SWAP operation takes the state to be
\begin{equation}
    \label{eq:disentangled_structure}
    \begin{tikzpicture}[scale=1]
        \clip (-0.6,0.65) rectangle + (2.2,-1.3);
        \draw[blue] (0,0) .. controls (-0.4,0.8) and (0.4,0.8) .. (0,0);
        \draw[blue] (1,0) .. controls (2,-0.5) and (3,-0.5) .. (4,0);
        \draw[blue] (1,0) .. controls (2,-0.8) and (3,-0.8) .. (4,0);
        \filldraw[black] (0,0) circle (2pt);
        \filldraw[black] (1,0) circle (2pt);
    \end{tikzpicture}
    \qquad \quad 
    \begin{tikzpicture}[scale=1]
        \draw[gray] (0,00)  -- (0,+1.3);
    \end{tikzpicture} 
    \quad\qquad\begin{tikzpicture}[scale=1]
        \clip (-0.6,0.65) rectangle + (2.2,-1.3);
        \draw[blue] (1,0) .. controls (0.6,0.8) and (1.4,0.8) .. (1,0);
        \draw[blue] (0,0) .. controls (-1,-0.5) and (-2,-0.5) .. (-3,0);
        \draw[blue] (0,0) .. controls (-1,-0.8) and (-2,-0.8) .. (-3,0);
        \filldraw[black] (0,0) circle (2pt);
        \filldraw[black] (1,0) circle (2pt);
    \end{tikzpicture}
\end{equation}
therefore reducing the entanglement by one unit across the bond.

Now, let us consider the following symmetric local entanglement structure:
\begin{equation}
    \label{eq:disentangled_structure_1}
    \begin{tikzpicture}[scale=1]
        \node[] at (0.,0.3) {$g_1$};
        \node[] at (0.,-0.7) {$g_2$};
    \end{tikzpicture}
    \begin{tikzpicture}[scale=1]
        \clip (-0.6,0.7) rectangle + (2.2,-1.5);
        \draw[blue] (0,0) .. controls (1,0.8) and (2,0.8) .. (3,0);
        \draw[blue] (0,0) .. controls (-1,0.8) and (-2,0.8) .. (-3,0);
        \draw[blue] (1,0) .. controls (0,-0.8) and (-1,-0.8) .. (-2,0);
        \draw[blue] (1,0) .. controls (2,-0.8) and (3,-0.8) .. (4,0);
        \filldraw[black] (0,0) circle (2pt);
        \filldraw[black] (1,0) circle (2pt);
    \end{tikzpicture}
    \begin{tikzpicture}[scale=1]
        \node[] at (0.,0.) {};
        \node[] at (0, 1.3) {$g_4$};
        \node[] at (0.,0.25) {$g_3$};
    \end{tikzpicture}
\end{equation}
There are 4 stabilizers that are relevant in such a case, and they have the following structure:
\begin{align*}
    g_1 &= \dots A_1\mathds{1}_2, \\
    g_2 &= \dots B_1C_2, \\
    g_3 &= \phantom{\dots} \mathds{1}_1F_2 \dots, \\
    g_4 &= \phantom{\dots} E_1D_2\dots,
\end{align*}
where the dots indicate that there is at least another site with nontrivial content. All the endpoints must be a Pauli matrix, while the rest ($B$ and $F$) can be either a Pauli matrix or the identity. We omit the tensor product in between operators for clarity of notation. Since all stabilizers must commute, we can observe several properties. First, $[A,E]=0$ and $[C,F]=0$, which implies that $A=E$ and $C=F$. Therefore, we write 
\begin{align*}
    g_1 &= \dots A_1\mathds{1}_2 \\
    g_2 &= \dots B_1C_2 \\
    g_3 &= \phantom{\dots} \,\mathds{1}_1C_2 \dots \\
    g_4 &= \phantom{\dots} A_1D_2\dots
\end{align*}
From the commutation of the strings, we also need that $[B_1C_2,A_1D_2]=0$. This means that either (a) $[C,D]=[B,A]=0$ or (b) $\{C,D\}=\{B,A\}=0$. The objective is to find a unitary $U$ such that $UA_1\mathds{1}_2U^\dagger=\tilde{A}_1\mathds{1}_2$ and $UB_1C_2U^\dagger=\tilde{B}_1\mathds{1}_2$, with $\{\tilde{A},\tilde{B}\}=0$. Since unitaries cannot change the commutation relations, we observe that case (a) cannot be disentangled. 

In case (b), we have that $\{A_1\mathds{1}_2, B_1C_2\}=0$, so the objective form can be obtained. To do so, we define a unitary that transforms the Pauli strings in the following way:
\begin{align*}
    UA_1U^\dagger &= A_1,\\
    UB_1U^\dagger &= B_1C_2,\\
    UC_2U^\dagger &= C_2,\\
    UD_2U^\dagger &= A_1D_2.
\end{align*}
This transformation preserves all the commutation relations, so it is a valid unitary. Observe that the unitary is completely fixed by choosing the Pauli matrices that are left invariant, $A_1$ and $C_2$. Each of them can take 3 values, so in total there are 9 such unitaries, which we denote by $\mathcal{U}_d=\{U_1,..., U_9\}$. These are all the unitaries required to disentangle states of the form~\eqref{eq:disentangled_structure_1}.

Next, we consider the entangled structure
\begin{equation}
    \label{eq:disentangled_structure_2}
    \begin{tikzpicture}[scale=1]
        \node[] at (0,0) {};
        \node[] at (0.,0.2) {$g_2$};
        \node[] at (0.,0.95) {$g_1$};
    \end{tikzpicture}
    \begin{tikzpicture}[scale=1]
        \clip (-0.6,0.5) rectangle + (2.2,-1.2);
        \draw[blue] (1,0) .. controls (0,0.5) and (-1,0.5) .. (-2,0);
        \draw[blue] (1,0) .. controls (0,-0.5) and (-1,-0.5) .. (-2,0);
        \draw[blue] (0,0) .. controls (1,-0.5) and (2,-0.5) .. (3,0);
        \draw[blue] (0,0) .. controls (1,0.5) and (2,0.5) .. (3,0);
        \filldraw[black] (0,0) circle (2pt);
        \filldraw[black] (1,0) circle (2pt);
    \end{tikzpicture}
    \begin{tikzpicture}[scale=1]
        \node[] at (0,0) {};
        \node[] at (0.,0.2) {$g_3$};
        \node[] at (0.,0.95) {$g_4$};
    \end{tikzpicture}
\end{equation}
which is the only one in which entanglement can be reduced by 2 units. In this case, the relevant stabilizers take the form
\begin{align*}
    g_1 &= \dots A_1B_2, \\
    g_2 &= \dots C_1D_2, \\
    g_3 &= \phantom{\dots} E_1F_2 \dots, \\
    g_4 &= \phantom{\dots} G_1H_2\dots.
\end{align*}
As before, we are going to consider two different cases. When $[A_1B_2, C_1D_2]=0$, the entanglement entropy can be reduced only by one. This can be done by just applying any unitary in $\mathcal{U}_d$ such that $UA_1B_2U^\dagger = A_1$. Then, we will get $Ug_3U^\dagger=\tilde{E}_1\tilde{F}_2\dots$ and $Ug_4U^\dagger=\tilde{G}_1\tilde{H}_2\dots$. Since commutation relations are preserved, they must commute with $Ug_1U^\dagger=\dots A_1$, so $[A,\tilde{E}]=[A,\tilde{G}]=0$. This can only hold if $\tilde{E}$ and $\tilde{G}$ are equal to $A$ or to the identity. If one of them is equal to the identity, then we are already in the situation of Eq.~\ref{eq:disentangled_structure_1}. Otherwise, we have $\tilde{E}=\tilde{G}$, so we can multiply the two stabilizers to get a stabilizer with an identity in site 1, going back to the clipped gauge and getting the form of Eq.~\ref{eq:disentangled_structure_1}.

If $\{A_1B_2, C_1D_2\}=0$, then by construction $[A,C]=0$ (since $\{B,D\}=0$). This can happen only in two cases: either when $A=C$ or when one of the two is the identity. In both cases, applying a SWAP gate will reduce the entanglement by at least 1, leaving the stabilizer structure as the one in~\eqref{eq:disentangled_structure_1}, and this structure can be completely disentangled with the unitaries $\mathcal{U}_d$.

We have shown that with the set of 19 unitaries $\{\mathcal{U}_d, \text{SWAP}, \mathcal{U}_d\times \text{SWAP}\}$ we can disentangle any locally entangled bond of the forms~\eqref{eq:disentangled_structure}, \eqref{eq:disentangled_structure_1}, and \eqref{eq:disentangled_structure_2}. There are six other locally entangled structures. Four of them are the following:
\begin{equation}
    \begin{tikzpicture}[scale=0.8]
        \clip (-0.6,0.85) rectangle + (1.9,-1.7);
        \draw[blue] (0,0) .. controls (0.33,0.5) and (0.66,0.5) .. (1,0);
        \draw[blue] (0,0) .. controls (0.33,-0.5) and (0.66,-0.5) .. (1,0);
        \filldraw[black] (0,0) circle (2pt);
        \filldraw[black] (1,0) circle (2pt);
    \end{tikzpicture}
    \begin{tikzpicture}[scale=1]
        \draw[gray] (0,00)  -- (0,+1.3);
    \end{tikzpicture} 
    \begin{tikzpicture}[scale=0.8]
        \clip (-0.3,0.85) rectangle + (1.9,-1.7);
        \draw[blue] (0,0) .. controls (1,0.8) and (2,0.8) .. (3,0);
        \draw[blue] (1,0) .. controls (1.66,0.6) and (2.33,0.6) .. (3,0);
        \draw[blue] (0,0) .. controls (0.33,-0.5) and (0.66,-0.5) .. (1,0);
        \filldraw[black] (0,0) circle (2pt);
        \filldraw[black] (1,0) circle (2pt);
    \end{tikzpicture}
    \quad
    \begin{tikzpicture}[scale=1]
        \draw[gray] (0,00)  -- (0,+1.3);
    \end{tikzpicture} 
    \quad
    \begin{tikzpicture}[scale=0.8]
        \clip (-0.6,0.85) rectangle + (1.9,-1.7);
        \draw[blue] (1,0) .. controls (0,0.8) and (-1,0.8) .. (-2,0);
        \draw[blue] (0,0) .. controls (-0.33,0.6) and (-1.66,0.6) .. (-3,0);
        \draw[blue] (0,0) .. controls (0.33,-0.5) and (0.66,-0.5) .. (1,0);
        \filldraw[black] (0,0) circle (2pt);
        \filldraw[black] (1,0) circle (2pt);
    \end{tikzpicture}    
    \,
    \begin{tikzpicture}[scale=1]
        \draw[gray] (0,00)  -- (0,+1.3);
    \end{tikzpicture} 
    \quad
    \begin{tikzpicture}[scale=0.8]
        \clip (-0.6,0.85) rectangle + (2.2,-1.7);
        \draw[blue] (1,0) .. controls (0,0.8) and (-1,0.8) .. (-2,0);
        \draw[blue] (0,0) .. controls (1,0.8) and (2,0.8) .. (3,0);
        \draw[blue] (0,0) .. controls (0.33,-0.5) and (0.66,-0.5) .. (1,0);
        \filldraw[black] (0,0) circle (2pt);
        \filldraw[black] (1,0) circle (2pt);
    \end{tikzpicture}   
\end{equation}
All of these states have a stabilizer $g_1=A_1B_2$ that is completely contained in the bond. Thus, disentangling the state just consists of applying a unitary in $\mathcal{U}_d$ such that $UA_1B_2U^\dagger=A_1$ or $UA_1B_2U^\dagger=B_2$.

Finally, the two remaining entanglement structures are:
\begin{equation}
    \begin{tikzpicture}[scale=1]
        \clip (-0.6,0.65) rectangle + (2.2,-1.3);
        \draw[blue] (0,0) .. controls (1,0.8) and (2,0.8) .. (3,0);
        \draw[blue] (0,0) .. controls (-1,0.8) and (-2,0.8) .. (-3,0);
        \draw[blue] (1,0) .. controls (0,-0.5) and (-1,-0.5) .. (-2,0);
        \draw[blue] (1,0) .. controls (0,-0.8) and (-1,-0.8) .. (-2,0);
        \filldraw[black] (0,0) circle (2pt);
        \filldraw[black] (1,0) circle (2pt);
    \end{tikzpicture}
    \quad 
    \begin{tikzpicture}[scale=1]
        \draw[gray] (0,00)  -- (0,+1.3);
    \end{tikzpicture} 
    \quad
    \begin{tikzpicture}[scale=1]
        \clip (-0.6,0.65) rectangle + (2.2,-1.3);
        \draw[blue] (0,0) .. controls (1,-0.5) and (2,-0.5) .. (3,0);
        \draw[blue] (0,0) .. controls (1,-0.8) and (2,-0.8) .. (3,0);              \draw[blue] (1,0) .. controls (0,0.8) and (-1,0.8) .. (-2,0);
        \draw[blue] (1,0) .. controls (2,0.8) and (3,0.8) .. (4,0);
        \filldraw[black] (0,0) circle (2pt);
        \filldraw[black] (1,0) circle (2pt);
    \end{tikzpicture}
\end{equation}
These two can also be disentangled using the unitaries in $\mathcal{U}_d$, as one can check following a similar argument as for the structure~\eqref{eq:disentangled_structure_1}. Therefore, the set of 19 unitaries that we have found is enough to disentangle any possible bond.

Observe that we have constructed only a possible set of unitaries that can disentangle any given bond, but we could consider many others. For example, applying one-qubit rotations after the unitaries would not change the disentangling power of the list, but would give a whole new set of matrices.

\section{Comparison of Clifford disentanglers}
\label{appendix_clifford_disentanglers}

As mentioned in the main text, the Clifford disentangler has several options to maximally disentangle a bond since different Clifford gates reduce the entanglement by the same amount. Here, we compare different ways to choose the disentangling and show that they are all equivalent. This indicates that the average trajectory does not depend on which disentangling unitary is selected as long as it maximally disentangles  the bond. We compare the following methods: (i) Random sampling, which checks the action of all Clifford unitaries in the given bond and chooses randomly a gate among the ones that disentangle it maximally. (ii)  Reduced random sampling, which uses a reduced set of 19 Clifford unitaries that are enough to disentangle any possible bond and tries all of them out. Then, it chooses a random gate among the optimal ones. (iii) Ordered sampling, which uses the same subset of the previous method, but the chosen unitary is the first that is found to disentangle the bond maximally (so that not all unitaries are necessarily tried out).

\begin{figure}[t!]
    \centering
    \includegraphics{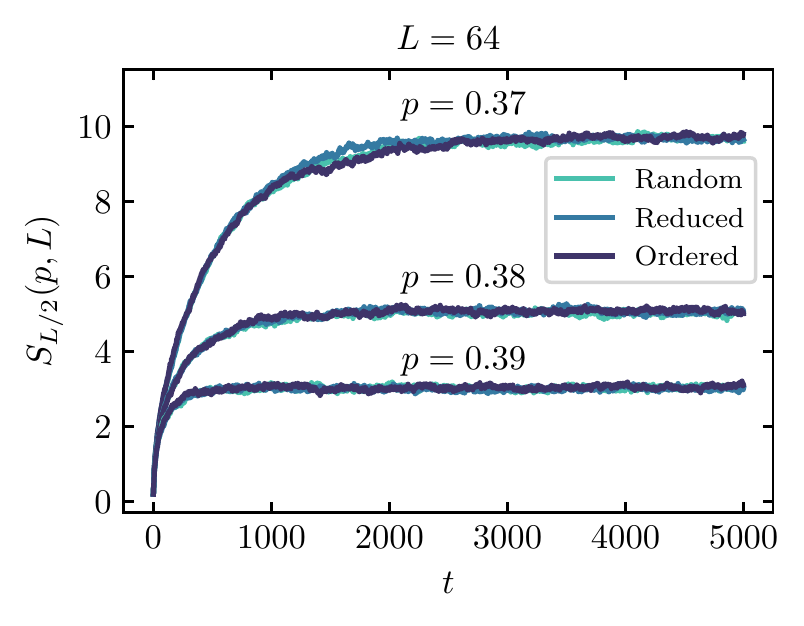}
    \caption{Comparison of the time evolution of the half-chain entanglement entropy for different Clifford disentanglers and different values of the disentangler probability for system size $L=64$. Circuit average with $10^2$ realizations.}
    \label{fig: clifford disentanglers}
\end{figure}

The time evolution of the half-chain entanglement entropy for system size $L=64$ is illustrated in Fig.~\ref{fig: clifford disentanglers}, where the three distinct sampling methods for different disentangling probabilities are compared. Each line represents the average of $10^3$ circuit realizations. Notably, all three methods yield identical results for the evolution of the entanglement entropy and the steady state value. Therefore, in our simulations, we use the ordered sampling method to disentangle bonds, since it requires the least number of tries to find the disentangling gate, allowing for faster simulations.

\section{Disentangling maximally entangled states}
\label{appendix_disentangling_Haar}

The task of the von Neumann disentangler is to find the optimal unitary in $U(4)$ that reduces maximally the bipartite entanglement on a given bond. This means that it has to perform a minimization over 9 continuous parameters. Here, we investigate how effective the disentangler is when trying to disentangle a maximally entangled state of $L$ qubits. To do so, we start with a product state and we evolve the system with a random circuit of depth $2L$, yielding a state with maximal entanglement. Then, the disentangler starts acting by minimizing the entanglement entropy in random bonds until the half-chain entanglement entropy is reduced below a certain threshold. We note that our disentangling procedure is similar to the Metropolis-like cooling algorithms introduced in Refs.~\cite{entanglement_cooling, entanglement_cooling2}. However, in that case, the disentangler acts with a random unitary and accepts it only when the entanglement entropy is reduced, avoiding the minimization step of our disentangler.

\begin{figure}[t!]
    \centering
    \includegraphics{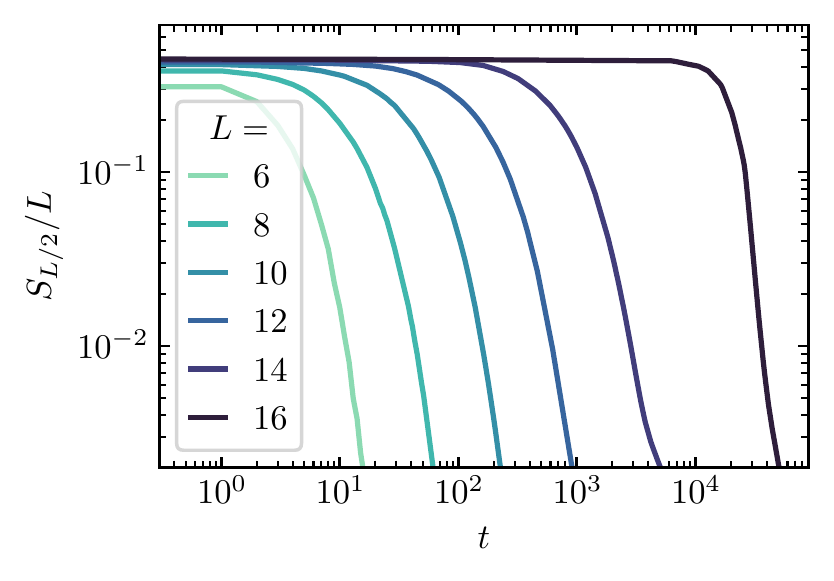}
    \caption{Time evolution of the half-chain entanglement entropy for a disentangling circuit, with von Neumann disentangler, starting with an $L$-qubit state generated by a depth $2L$ random circuit. Circuit averaged over $10^2$ realizations.}
    \label{fig: haar disentangler}
\end{figure}

Figure~\ref{fig: haar disentangler} shows the averaged time evolution of the half-chain entanglement entropy for the disentangling circuit. The time it takes to disentangle  grows exponentially with system size. When looking at the behavior of the $L=16$ line, we observe that the evolution gets stuck in the maximally entangled state for a long time before being able to effectively reduce the entanglement. We attribute this behavior to numerical instabilities, where the optimization fails to find the global minimum of the entanglement entropy. Furthermore, we remark that in contrast with the entanglement cooling algorithm in Refs.~\cite{entanglement_cooling, entanglement_cooling2}, our disentangler is able to disentangle the state given a sufficiently long time.

\section{Disentangling stabilizer states}

The absence of a phase transition in the Haar case is due to the creation of complicated entanglement structures that cannot be efficiently locally disentangled. Instead, entanglement in stabilizer states has a very simple structure, as we have shown in App.~\ref{appendix_clifford_gates_disentangling}, and can be locally reduced easily. In particular, as we are going to show next, any stabilizer state can be locally disentangled using a circuit with depth $\mathcal{O}(L)$. 

\begin{figure}[t!]

    \centering
    \includegraphics{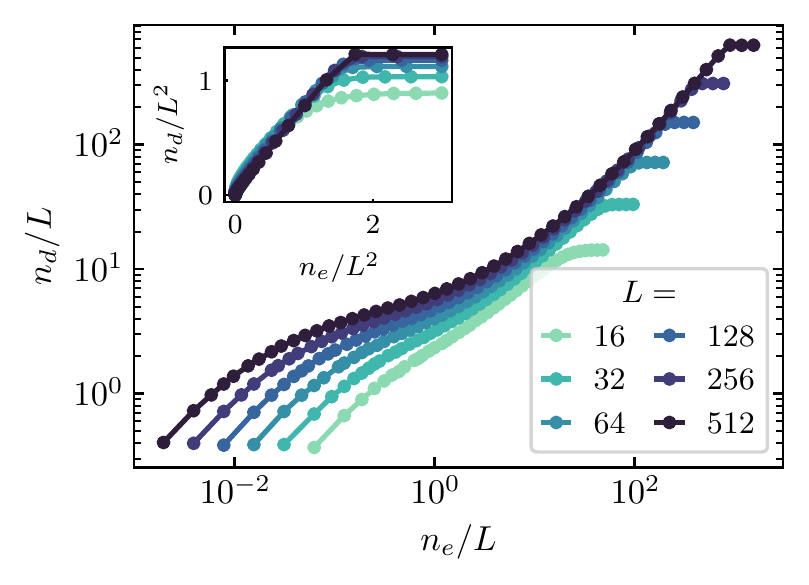}
    \caption{Average number of disentangling steps $n_d$ needed to completely disentangle a state created by a circuit of $n_e$ randomly placed two-qubit Clifford unitary. The inset shows the collapse for $n_d/L^2$ vs $n_e/L^2$.}
    \label{fig: disentangle circuit depth clifford}
\end{figure}

To study the disentangling complexity of Clifford circuits, we are going to consider the following setup: start with a product stabilizer state and evolve with a random circuit with $n_e$ gates. Then, run a disentangling circuit until entanglement is reduced to 0, and count how many two-qubit gates were required, $n_d(n_e)$. The numerical results are shown in Fig.~\ref{fig: disentangle circuit depth clifford}. We observe three different regimes. For $n_e<L$, there is a nonlinear regime in which reducing the entanglement just consists of finding the entangled bonds and applying the inverse Clifford unitary. For $n_e > L^2$, the number of disentangling gates becomes independent of $n_e$. This is the regime where the state is maximally entangled, and therefore adding more entangling gates does not change the required depth of the disentangling circuit. In such case, the number of gates required to completely disentangle is $n_d \propto L^2$, as shown in the inset of Fig.~\ref{fig: disentangle circuit depth clifford}. In the regime in between, $L<n_e<L^2$, the depth of the disentangling circuit grows linearly with the depth of the entangling circuit, with a system-size independent slope. These results contrast with the random Haar case shown in Fig.~\ref{fig: disentangle circuit depth}, where for $n_e>L$ the depth of the disentangling circuit becomes exponential with system size.

Note that, for $n_e<L$, the number of disentangling steps required is not the same as for generic circuits. This is caused by the finite probability that a random Clifford unitary does not entangle the system at all, which is not possible for random Haar gates.

\newpage
\end{appendix}

\bibliography{references}

\end{document}